\documentclass[11pt,oneside,final]{YiMSthesis}
\usepackage{epsfig,bm,epsf,float,rotating,CJK}  
\usepackage{bibunits}  
\usepackage{cite}  
\usepackage{threeparttable}  
\usepackage{amsmath}
\pdfoutput=1

\begin{document}

%

\newcommand{\deriv}[2]{\frac{d#1}{d#2}}
\newcommand{\derivc}[3]{\left. \frac{d#1}{d#2}\right|_{#3}}
\newcommand{\pd}[2]{\frac{\partial #1}{\partial #2}}
\newcommand{\pdc}[3]{\left. \frac{\partial #1}{\partial #2}\right|_{#3}}

\newcommand{\bra}[1]{\left\langle #1\right|}
\newcommand{\ket}[1]{\left|#1\right\rangle}
\newcommand{\braket}[2]{\left\langle #1 \left|#2\right.\right\rangle}
\newcommand{\braOket}[3]{\left\langle #1\left|#2\right|#3\right\rangle}

\def\cal#1{\mathcal{#1}}

\def\avg#1{\left< #1 \right>}
\def\abs#1{\left| #1 \right|}
\def\recip#1{\frac{1}{#1}}
\def\vhat#1{\hat{{\bf #1}}}
\def\smallfrac#1#2{{\textstyle\frac{#1}{#2}}}
\def\smallrecip#1{\smallfrac{1}{#1}}

\def\spshalf{{1\over{2}}}
\def\Orabi{\Omega_{\rm rabi}}
\def\btt#1{{\tt$\backslash$#1}}


\def\schrod{Schroedinger's Equation}
\def\helm{Helmholtz Equation}

\def\be{\begin{equation}}
\def\ee{\end{equation}}
\def\bea{\begin{eqnarray}}
\def\eea{\end{eqnarray}}
\def\bean{\begin{mathletters}\begin{eqnarray}}
\def\eean{\end{eqnarray}\end{mathletters}}

\newcommand{\tbox}[1]{\mbox{\tiny #1}}
\newcommand{\half}{\mbox{\small $\frac{1}{2}$}}
\newcommand{\pit}{\mbox{\small $\frac{\pi}{2}$}}
\newcommand{\sfrac}[1]{\mbox{\small $\frac{1}{#1}$}}
\newcommand{\mbf}[1]{{\mathbf #1}}
\def\text{\tbox}

\newcommand{\mV}{{\mathsf{V}}}
\newcommand{\mL}{{\mathsf{L}}}
\newcommand{\mA}{{\mathsf{A}}}
\newcommand{\lB}{\lambda_{\tbox{B}}}  
\newcommand{\ofr}{{(\mbf{r})}}       
\def\ofkr{(k;\mbf{r})}			
\def\ofks{(k;\mbf{s})}			
\newcommand{\ofs}{{(\mbf{s})}}       
\def\xt{\mbf{x}^{\tbox T}}		

\def\ce{\tilde{C}_{\tbox E}}		
\def\cew{\tilde{C}_{\tbox E}(\omega)}		
\def\ceqmw{\tilde{C}^{\tbox{qm}}_{\tbox E}(\omega)}	
\def\cewqm{\tilde{C}^{\tbox{qm}}_{\tbox E}}	
\def\ceqm{C^{\tbox{qm}}_{\tbox E}}	
\def\cw{\tilde{C}(\omega)}		
\def\cfw{\tilde{C}_{\cal F}(\omega)}		

\def\tcl{\tau_{\tbox{cl}}}		
\def\tcol{\tau_{\tbox{col}}}		
\def\terg{t_{\tbox{erg}}}		
\def\tbl{\tau_{\tbox{bl}}}		
\def\theis{t_{\tbox{H}}}		

\def\area{\mathsf{A}_D}			
\def\ve{\nu_{\tbox{E}}}			
\def\vewna{\nu_E^{\tbox{WNA}}}		

\def\dxcqm{\delta x^{\tbox{qm}}_{\tbox c}}	

\newcommand{\rop}{\hat{\mbf{r}}}	
\newcommand{\pop}{\hat{\mbf{p}}}

\newcommand{\sint}{\oint \! d\mbf{s} \,} 
\def\gint{\oint_\Gamma \!\! d\mbf{s} \,} 
\newcommand{\lint}{\oint \! ds \,}	
\def\infint{\int_{-\infty}^{\infty} \!\!}	
\def\dn{\partial_n}				
\def\aswapb{a^*\!{\leftrightarrow}b}		
\def\eps{\varepsilon}				

\def\dhdxt{\partial {\cal H} / \partial x}
\def\dhdx{\pd{\cal H}{x}}
\def\dhdxnm{\left( \pd{\cal H}{x} \right)_{\!nm}}
\def\dhdxnmsq{\left| \left( \pd{\cal H}{x} \right)_{\!nm} \right| ^2}

\def\bcs{\stackrel{\tbox{BCs}}{\longrightarrow}}	

\def\wx{\omega_x}
\def\wy{\omega_y}
\newcommand{\ofro}{({\bf r_0})}
\def\Eb{E_{\rm blue,rms}}
\def\Er{E_{\rm red,rms}}
\def\Es2{E_{0,{\rm sat}}^2}
\def\sb{s_{\rm blue}}
\def\sr{s_{\rm red}}

\def\ie{{\it i.e.\ }}
\def\eg{{\it e.g.\ }}
\newcommand{\etal}{{\it et al.\ }}
\newcommand{\ibid}{{\it ibid.\ }}

\def\gap{\hspace{0.2in}}

%

\newcounter{eqletter}
\def\mathletters{%
\setcounter{eqletter}{0}%
\addtocounter{equation}{1}
\edef\curreqno{\arabic{equation}}
\edef\@currentlabel{\theequation}
\def\theequation{%
\addtocounter{eqletter}{1}\thechapter.\curreqno\alph{eqletter}%
}%
}
\def\endmathletters{\setcounter{equation}{\curreqno}}


\newcommand{\bk}{{\bf k}}
\def\kf{k_{\text F}}
\newcommand{\br}{{\bf r}}
\newcommand{\TL}{{\text{(L)}}}
\newcommand{\TR}{{\text{(R)}}}
\newcommand{\TLR}{{\text{L,R}}}
\newcommand{\VSD}{V_{\text{SD}}}
\newcommand{\GT}{\Gamma_{\text{T}}}
\newcommand{\DEL}{\mbox{\boldmath $\nabla$}}
\def\lf{\lambda_{\text F}}
\def\st{\sigma_{\text T}}
\def\stlr{\sigma_{\text T}^{\text{L$\rightarrow$R}}}
\def\strl{\sigma_{\text T}^{\text{R$\rightarrow$L}}}
\def\aeff{a_{\text{eff}}}
\def\aaeff{A_{\text{eff}}}
\def\gat{G_{\text{atom}}}
\newcommand{\LB}{Landauer-B\"{u}ttiker}

%
%


\hsp


\begin{figure}
	\vskip-1.53in
  \hskip-1.49in
		\scalebox{1.10}[1.10]{\includegraphics{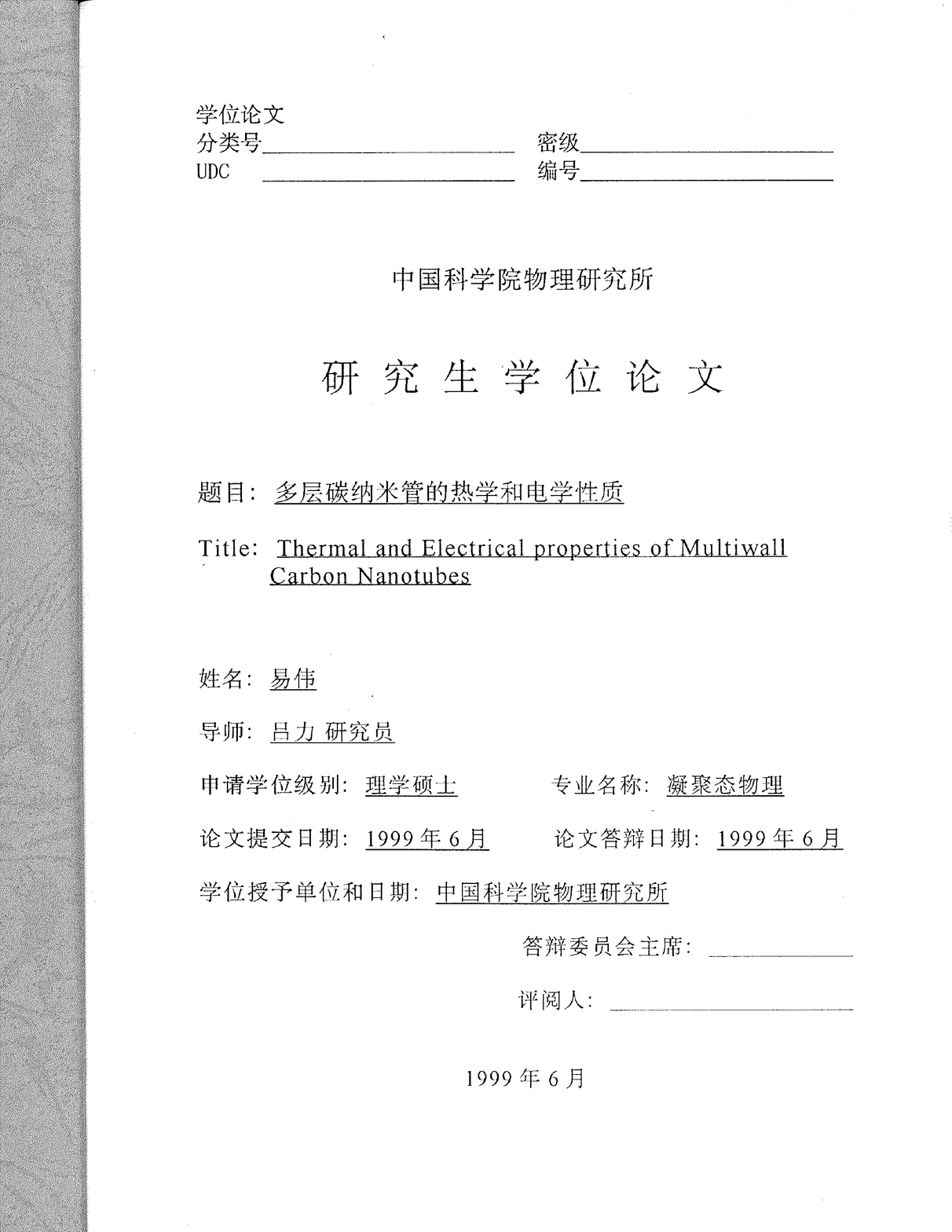}}
	\label{cover}
\end{figure}

\title{Thermal and Electrical Properties of Multiwall Carbon Nanotubes}
\author{Wei Yi}
\degreemonth{June} 
\degreeyear{1999}
\degree{Master of Science}
\field{Condensed Matter Physics}
\department{Institute of Physics}

\maketitle
\copyrightpage

\newpage
\begin{CJKAbstract}
\begin{CJK*}{GB}{gbsn}
\CJKtilde
\vspace*{-.8in}
  {\centerline{\bf {\LARGE 摘\  \ 要}}}
\vspace*{.2in}
\indent

　　　本文利用有机气体催化热解法制备的毫米长的取向多层碳纳米
管，对它的热学和电学性质进行了详细研究。
	本文共分六章。第一章为绪论，扼要概述了碳纳米管的发现、结
构、性质（包括理论研究和实验研究）。第二章概述了有机气体催化热
解法制备毫米长取向多层碳纳米管的制备工艺和表征。
	在第三章中，利用一种新的自加热3ω方法测量了碳纳米管
的热导、比热和热扩散系数。发现直径为几十纳米的多层碳纳米管的比热
在全部测量温区（10－300K）随温度呈很好的线性关系。该结果显示多层
碳纳米管的层间耦合比石墨的弱得多，因此在热学性质上多层碳纳米管
可以看作弱耦合的二维单层管的集合。热导率在高于120K的温区随温度
大致呈线性关系，而低于120K则呈二次方关系。热导率的幅值较低，反
映出热解法制备的多层碳纳米管可能存在较多的缺陷。
	在第四章中，首次用四点法测量了多层碳纳米管和正常金属之间
的单隧道结的低温隧道谱。观察到隧道态密度的强烈“零偏压抑制” 现
象，表现为隧道结微分电导随偏压和温度的幂次率关系。利用环境量子涨
落理论对微分电导随偏压和温度的关系进行了定量拟合，发现幂次率指数
是个由碳纳米管内库仑作用强度决定的普适常数。在两个有较小的结阻抗
的样品中，发现在温度低于1K且偏压小于1mV的区域，隧道电导出现呈
Fano线型的非对称共振峰，该现象可以用开放的量子点中的Kondo物理
来解释。
	在第五章中，对多层碳纳米管的热电势和纵向磁阻进行了研究。
利用比较法测量了碳纳米管的热电势。发现碳纳米管有比较大的正热电
势，在低温区（小于大约100K）随温度呈很好的金属性的线性关系。这
样一个大的正热电势与石墨完全不同，可能意味着金属性碳纳米管中由
于缺陷等可能的机制导致电子－空穴对称性的破缺。在20mK的极低温下
测量了碳纳米管的纵向磁阻。发现电阻随磁场增加呈现很好的周期性振
荡行为。如果只考虑最外层管壁对电导的贡献，则振荡周期与具有h/2e
周期的Altshuler-Aronov-Spivak（AAS）效应一致。该结果清晰地反映
出低温下多层碳纳米管中的量子相干行为，利用理论拟合得出的相位相
干长度与最外层管子的周长相当。
	第六章为结论。

\vspace{1.0in}

关键词：石墨，碳纳米管，比热，热导，3ω方法，隧道谱，库仑
作用，Fano共振，热电势，磁阻，AAS效应，量子相干

\end{CJK*}
\end{CJKAbstract}

\newpage
\begin{abstract}

In this dissertation, extensive researches on the thermal and electrical properties of millimeter-long aligned multiwall carbon nanotubes (MWNTs) prepared by thermal decomposition of hydrocarbons have been presented.

The thesis consists of six chapters. Chapter 1 is an introduction, in which the discovery, structure and physical properties (including theoretical and experimental) are outlined. In Chapter 2, the preparation techniques and the characterizations of the carbon nanotubes are described.

In Chapter 3, by using a novel self-heating 3$\omega$ method, the specific heat, diffusivity coefficient and thermal conductivity of MWNTs are for the first time reported. MWNTs of a few tens nm diameter is found to demonstrate a strikingly linear temperature-dependent specific heat over the entire temperature range measured (10--300 K). The results indicate that inter-wall coupling in MWNTs is rather weak compared with in its parent form, graphite, so that one can treat a MWNT as a few decoupled two-dimensional (2D) single wall tubules. The thermal conductivity $\kappa$ of MWNTs has a roughly linear temperature dependence above $\sim$120 K, below $\sim$120 K, $\kappa$ follows almost a quadratic $T$-dependence. The room-temperature amplitude of $\kappa$ is found to be low, indicating the existence of substantial amount of defects in the MWNTs prepared by chemical-vapor-deposition method.

In Chapter 4, the first true four-probe measurement of tunneling spectroscopy for single tunnel
junctions formed between MWNTs and a normal metal is presented. The intrinsic Coulomb interactions in the MWNTs give rise to a strong zero-bias suppression of tunneling density of states (TDOS) that can be fitted numerically with the environmental quantum-fluctuation (EQF) theory. At low temperatures, an asymmetric conductance anomaly near zero bias is observed, which is interpreted as Fano resonance in the strong tunneling regime.

In Chapter 5, the thermoelectric power (TEP) and longitudinal magnetoresistance (MR) of MWNTs are measured. MWNTs have a moderately large positive TEP. At low temperatures, TEP has a metallic-like linear temperature dependence. The results give strong evidence that the electron-hole symmetry in metallic nanotubes is broken. The longitudinal MR of MWNTs is measured at 20 mK. Periodic oscillations are observed when a longitudinal magnetic field is applied. If only the outermost graphene wall of MWNTs contributes to electrical conductance, the period of oscillation agrees well with the period $h/2e$ of Altshuler-Aronov-Spivak (AAS) effect. The present results clearly indicate that quantum-interference dominate the resistance of MWNTs at low temperatures.

Chapter 6 presents the main conclusions of this thesis.

\vspace{1.0in}


\end{abstract}

\newpage
\addcontentsline{toc}{section}{Table of Contents}
\tableofcontents


\begin{citations}

\vspace{0.8in}

\ssp
\noindent
\begin{quote}
\parindent 6pt
{1. \underline{W. Yi}, L. Lu, Zhang Dian-lin, Z. W. Pan, and S. S. Xie, {\it Linear specific heat of carbon nanotubes}, {\it Phys. Rev.} 
{\bf B 59}, R9015 (1999).

\vspace{0.5in}

2. L. Lu, \underline{W. Yi}, and Zhang Dian-lin, {\it A 3$\omega$ method for both specific heat and thermal conductivity measurement}, in preparation. (Published as: L. Lu, W. Yi, and D. L. Zhang, {\it 3-omega method for specific heat and thermal conductivity measurements}, Rev. Sci. Instrum. {\bf 72}, 2996 (2001).)

\vspace{0.5in}

3. \underline{W. Yi}, L. Lu, Zhang Dian-lin, Z. W. Pan, and S. S. Xie, {\it Electron-electron correlations in multiwall carbon nanotubes detected by tunneling spectroscopy}, in preparation. (Published as: W. Yi, L. Lu, H. Hu, Z. W. Pan, and S. S. Xie, {\it Tunneling into Multiwalled Carbon Nanotubes: Coulomb Blockade and the Fano Resonance}, Phys. Rev. Lett. {\bf 91}, 076801 (2003).)}
\end{quote}
\end{citations}

\dedication

\begin{quote}
\hsp
\em
\raggedleft

\large{Dedicated to my parents}\\

\end{quote}

\newpage

\startarabicpagination


\chapter{Introduction}

\begin{bibunit}[unsrt]

\section{Discovery of carbon nanotubes}

Carbon is really a special element in nature, whose chemical versatility makes it the central agent in most organic and inorganic chemistry as well as the fundamental constituent of all biological systems.

Until recently, pure carbon was only known to exist in two forms: diamond and graphite. In 1985, Harold Kroto, Robert Curl and Richard E. Smally discovered a new form of carbon--the fullerenes, which are molecules of pure carbon atoms bonded together forming geometrically regular structures \cite{Kroto}. The best-known of these is $C_{60}$, which has precisely the same geometry as a soccer ball, with one carbon atom at each point where the seams intersect. Because its structure is similar to that of the geodesic dome developed by the American architect, Buckminster Fuller, so this new molecule was named buckminsterfullerene, or buckyball. The discovery of fullerenes has added an important milestone in carbon research.

Most importantly, research has proven the possibilities for producing a whole new range of carbon structures of various sizes, shapes and dimensionalities. According to Euller theorem, infinite kinds of closed cage structure can be made by add  a minimum number of 12 pentagons to carbon honeycomb lattices. Soon later people discovered various such structures, such as $C_{70}$, $C_{76}$, $C_{84}$, $C_{96}$, etc. Naturally, the possibility of existance of tube-sized fullerene molecules, i.e., carbon nanotubes, has also been proposed.

In 1991, S. Iijima firstly observed the existence of carbon nanotube while studying the surface of carbon electrodes used in an electric arc-discharge appartus which had been used to make fullerenes \cite{Iijima}. He found needle-like carbon whiskers with diameter of 4--30 nm and length of 1 $\mu$m. These carbon whiskers are hollow tubes made of concentric clinders of carbon honeycomb lattices. The distance between adjacent cylinders is similar to inter-layer distance of graphite, i.e., $d\sim0.34$ nm. The discovery of carbon nanotubes stimulated tremendous interest in scientific community immediately. The subsequent success of the large-scale synthesis of nanotubes by T. W. Ebbesen and P. M. Ajayan of NEC opened the door for their widespread study. In 1993, Iijima \cite{Iijima2} and D. S. Bethune \cite{Bethune} firstly observed single-wall carbon nanotubes (SWNTs) with diameter only $\sim 1$ nm. The relative simplicity of structure of SWNTs made them suitable for theoretical calculations. By 1996, A. Thess successly synthesized large quantities of high-quality SWNTs using pulsed laser evaporation technique \cite{Thess}. Taking advantage of Scanning Probe Microscopy techniques obtainable nowadays, experimental explorations on the properties of individual SWNT made rapid progress, which has found extremely good agreement with theoretical predications.

\section{Structure}

\begin{figure}
	\centering
		\includegraphics[width=0.90\textwidth]{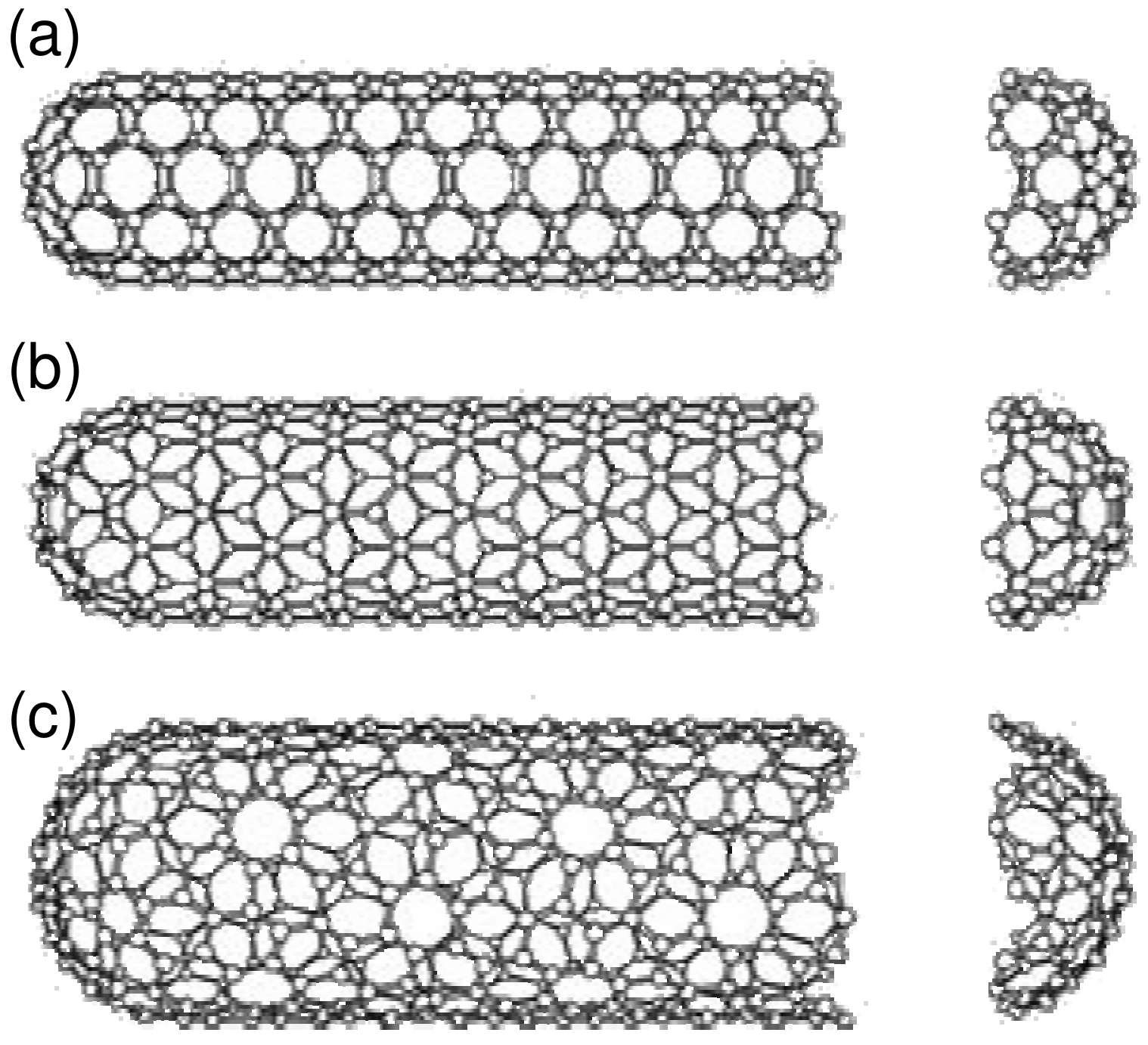}
	\caption{Schematic of the formation of carbon nanotubes by wrapping up graphene sheet. Different means of wrapping produce different tube structures: (a) armchair, (b) zigzag, (c) chiral. From Ref.\ \protect{\cite{DresselhausBook}}}
	\label{formation}
\end{figure}

Carbon nanotubes are made up of one or more wrapped-up seemless cylinders of carbon honeycomb lattices. The theoretically smallest nanotubes have a diameter equal to that of $C_{60}$ ($d=0.7$ nm), which could be thought as two hemispheres of $C_{60}$ connected by a single wrapped-up graphite sheet. If a $C_{60}$ molecule is separated perpendicular to the quintic symmetry axis, the derived tube is called `armchair' (Fig.\ \ref{formation} a); if it is separated perpendicular to the cubic symmetry axis, the derived tube is called `zigzag' (Fig.\ \ref{formation} b). We can see that the orientation of carbon hexagonal rings is either parallel or vertical to the tube axis for these two kinds of tubes. Actually for most of the nanotubes, the orientation of hexagonal rings is neither parallel nor vertical to the tube axis, but has some fixed helical angel. Such nanotubes have chirality (Fig.\ \ref{formation} c).

\begin{figure}
	\centering
		\includegraphics[width=0.90\textwidth]{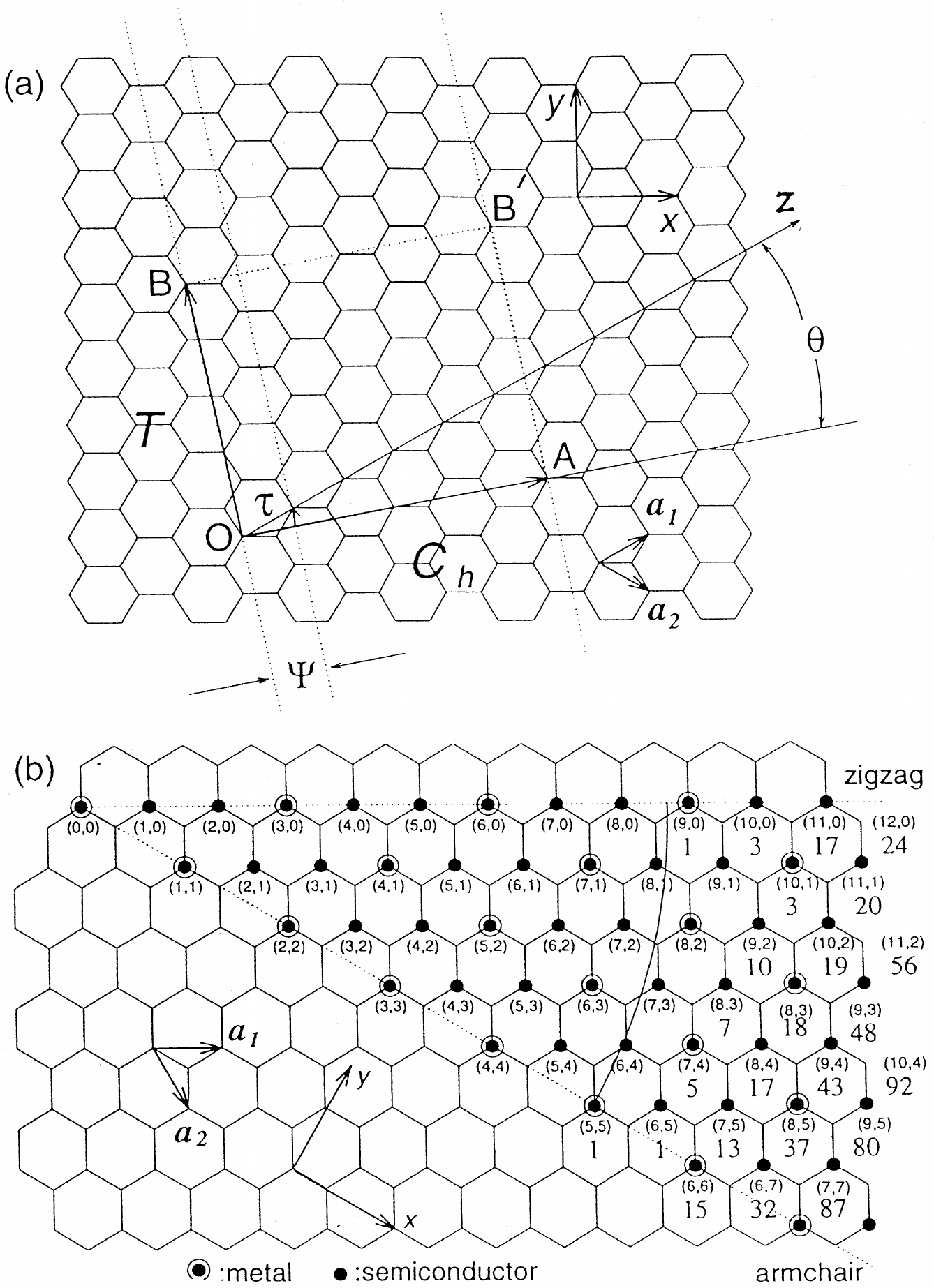}
	\caption{Schematic of the indexing method developed by Hamada et al. for single shells of carbon nanotubes. For example, a (6,3) nanotube is made by folding the two-dimensional hexagonal lattice shown and matching the lattice point (6,3) with the origin. The directions along which lattice points would make armchair and zigzag nanotubes are indicated. From Ref.\ \protect{\cite{DresselhausBook}}}
	\label{indexing}
\end{figure}

\vspace*{1cm}
In 1992, Hamada et al. developed  an indexing method to represent the structure for single shell of carbon nanotubes \cite{Hamada-Chp1}. As shown in Fig.\ \ref{indexing} a, cabon atoms are located at the corners of honeycomb lattice. The two primitive lattice vectors are $\vec{a}_{1,2}=(a/2)(\pm 1, \sqrt{3})$, where $a=\sqrt{3}d$, with $d=0.1421$ nm is the nearest-neighbour carbon separation. All the possible structures of carbon nanotubes can be represented by a pair of integers (n, m). A (n, m) nanotube is made by folding the two-dimensional hexagonal lattice shown and matching the lattice point (n, m) with the origin, with a wrapping vector $\vec{C}=n\vec{a}_{1}+m\vec{a}_{2}$. If we define the wrapping orientation of zigzag nanotube as the reference axis, the separation angle between the wrapping vector $\vec{C}$ and the reference axis is the helical angle $\theta$. All nanotubes are uniquely defined by the helical angle $\theta$ and the distance $\vec{C}$ which defines the tube size. Because of the symmetry of graphite lattice, $\theta$ is chosen as $0<\theta<30^{\circ}$ to avoid duplicity. The three kinds of nanotubes have different helical angle $\theta$. For armchair tubes, $n=m$, $\theta=30^{\circ}$. The zigzag tubes correspond to $m=0$ and $\theta=0$. When $n\neq m$ and not equal to zero, they are helical or chiral tubes with $0<\theta<30^{\circ}$.

For a (n, m) carbon nanotube, the diameter $d$ and helical angle $\theta$ are given by

\begin{equation}
d=\frac{\sqrt{3}}{\pi}a_{c-c}(n^2+nm+m^2)^{\frac{1}{2}}
\end{equation}
\begin{equation}
\theta=\arctan{\sqrt{3}/(m+2n)}
\end{equation}

\section{Physical properties, theoretical}

\begin{figure}
	\centering
		\includegraphics[width=0.40\textwidth,angle=90]{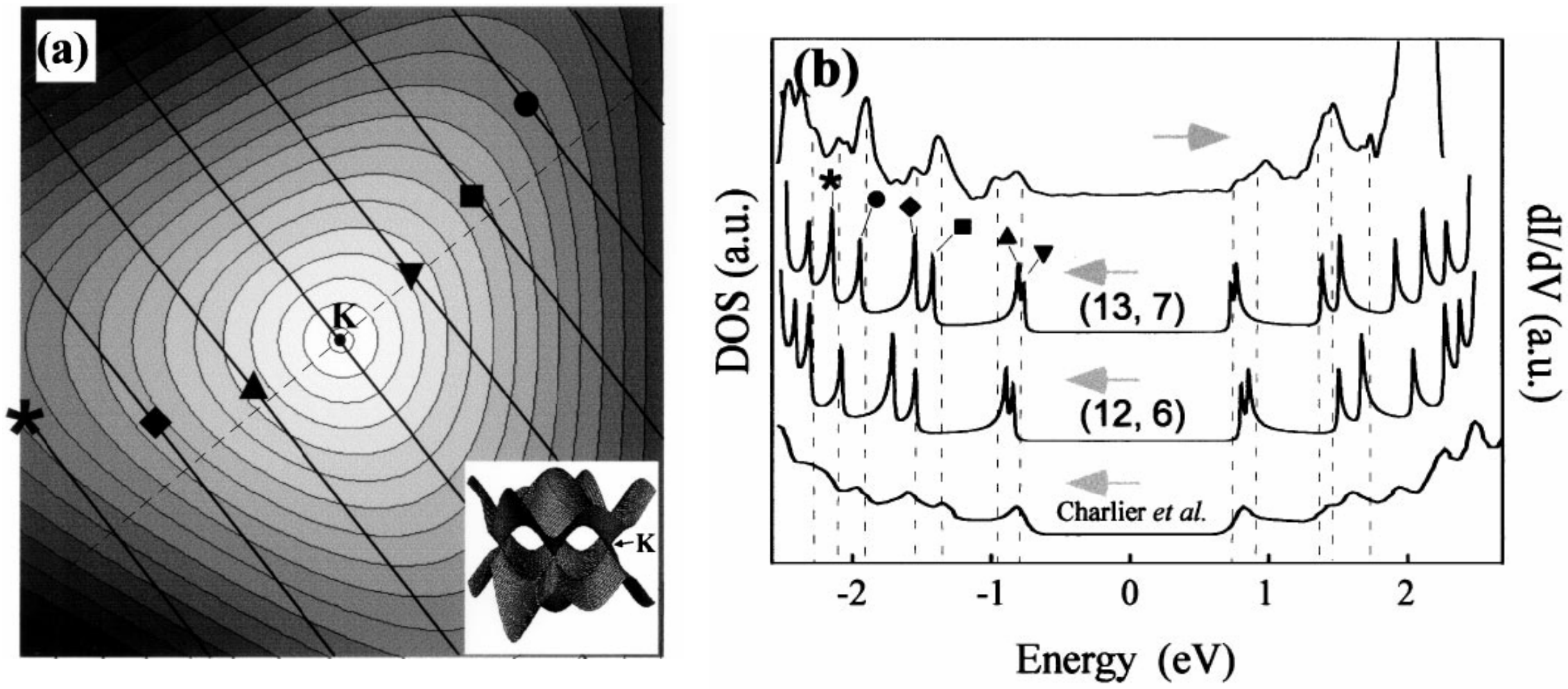}
	\caption{(a) Energy dispersion of the $\pi$-band of graphene sheet near the $K$. The solid lines correspond to 1D bands obtained by the zone-folding. Symbols are located at the positions where Van Hove singularities occur in these 1D bands. The inset depicts a three dimensional view of graphene $\pi/\pi^{*}$ bands. (b) Comparison of the electron density of states obtained from scanning tunneling microscopy experiment (upper curve) and tight-binding calculations (lower curves). From Ref.\ \protect{\cite{Kim}}}
	\label{dispersion}
\end{figure}

Carbon nanotubes as a novel quasi one-dimensional material has stimulated great interest. From theoretical point of view, singlewall carbon nanotubes is ideal for study because of their relative simplicity. A straight-forward tight-binding calculation is based on the zone-folding picture, which deduces the band structure of carbon nanotubes from their parent form, graphene sheet, with periodic boundary conditions around circumference of nanotubes \cite{Hamada-Chp1,Mintmire,Mintmire-2,RSaito,RSaito2}. For a graphene sheet, each atom has one conduction electron in the $2p_z$ state. The first Brillouin zone (BZ) is a hexagon whose sides are distant $1/3a_{c-c}$ from its center. It is easily shown that this zone contains one electron per atom, so graphite is a half-filling semiconductor with zero gap, or a semimetal (shown in Fig.\ \ref{dispersion}). Fermi surface reduces into six points at the corner of hexagonal BZ. Near Fermi level, the dispersion relation is nearly linear \cite{Wallace-Chp1}.

\begin{figure}
	\centering
		\includegraphics[width=0.6\textwidth]{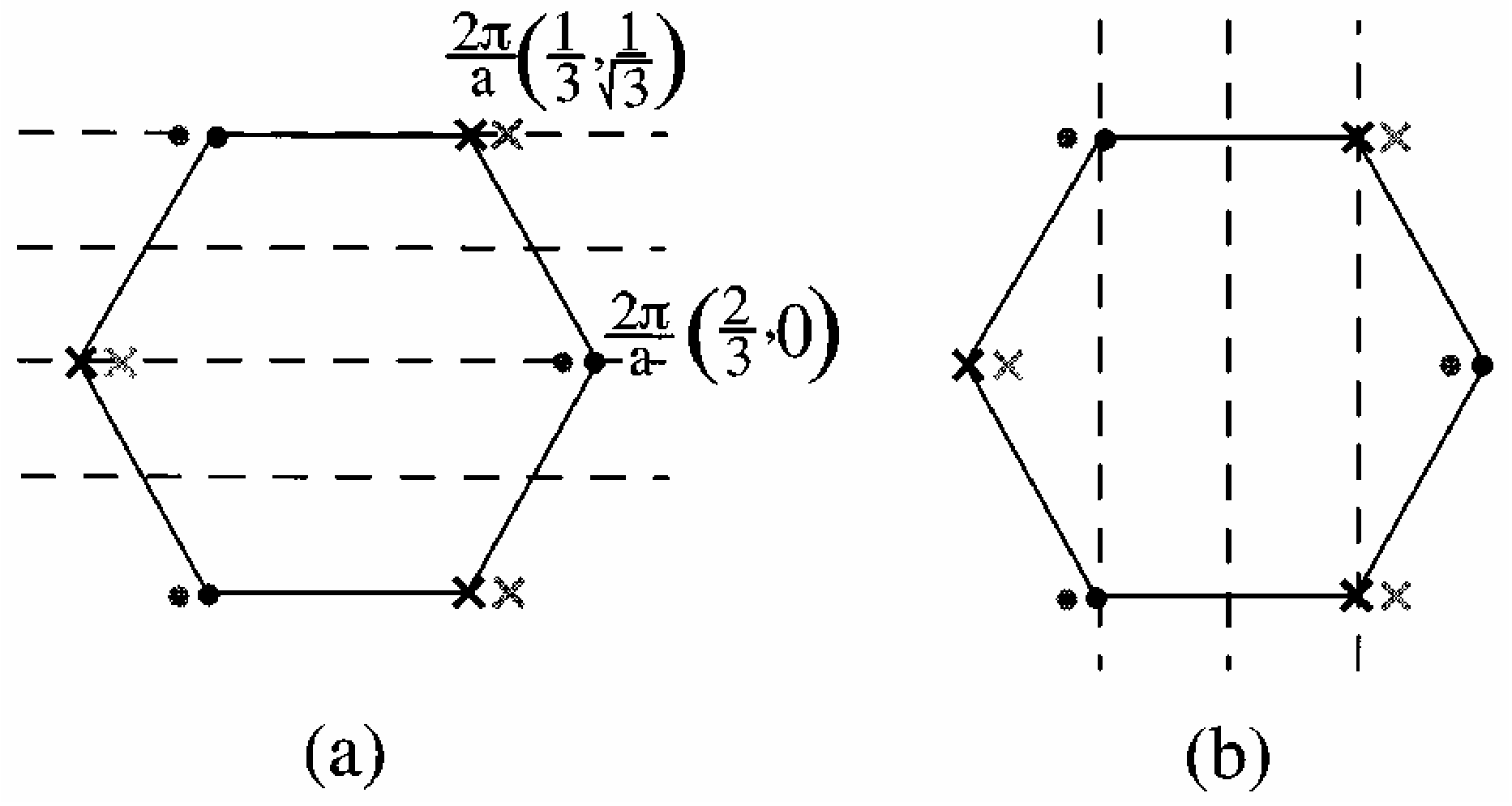}
	\caption{Fermi points in the Brillouin zone of carbon nanotubes. Dark circles and crosses indicate the locations of the gapless points without curvature-induced symmetry breaking, while gray symbols schematically indicate the shifted positions for curvature effects. Dashed lines cut the zone at a discrete set of allowed transverse momenta in (a) the armchair tube and (b) the zig-zag tube (here with N=3). From Ref.\ \protect{\cite{Balents}}}
	\label{fermi}
\end{figure}

The periodic boundary conditions around circumference of nanotube make the crystal momentum transverse to the tube axis quantized. The allowed electronic states are restricted to those in the graphene BZ that satisfy \cite{Mintmire-2} 

\begin{equation}
\vec{k}\cdot\vec{C}=2\pi m
\end{equation}

with $m$ an integer (Fig.\ \ref{fermi}). 
In terms of the two-dimensional BZ of graphene, the allowed states will lie along parallel lines separated by a spacing of $2\pi/\vert\vec{C}\vert$. Metallic behavior happens whenever the gapless Fermi points lie on the allowed transverse quantized wave vectors. For (n, n) armchair nanotubes, the gapless modes are present at $k_{y}=0$, resulting in metallic behavior independent of n. For zigzag or chiral tubes with (n, m), metallic behavior happens only if the vectors of Fermi points satisfy Eq. (1.3), leading to the condition $n-m=3q$, where $q$ is an integer. For the nonmetallic case, the smallest band gap for the nanotube will occur at the nearest allowed point to $\vec{K}$, the Fermi wavevector, with amplitude given by tight-binding H$\ddot{u}$kel model \cite{Mintmire-2}

\begin{equation}
E_g=V_{pp\pi}d/R
\end{equation}
where $V_{pp\pi}$ is hopping matrix element of the $\pi$-bands, $R$ is the nanotube radius($R=\vert\vec{C}\vert/2\pi$).

This simplified model does not consider the symmetry breaking introduced by curvature effects. Due to curvature effects, the Fermi points will be shifted slightly along $k_x$, but leaves the armchair tube gapless (Fig.\ \ref{fermi}a). For chiral tubes which satisfie $n-m=3q$, Fermi points are shifted away from the allowed quantized wave vectors slightly by order of $1/R^2$. Thus chiral tubes are quasimetallic with second-order small band gaps (Fig.\ \ref{fermi}b) \cite{Kane-Chp1,Balents}.

\begin{figure}
	\centering
		\includegraphics[width=0.6\textwidth]{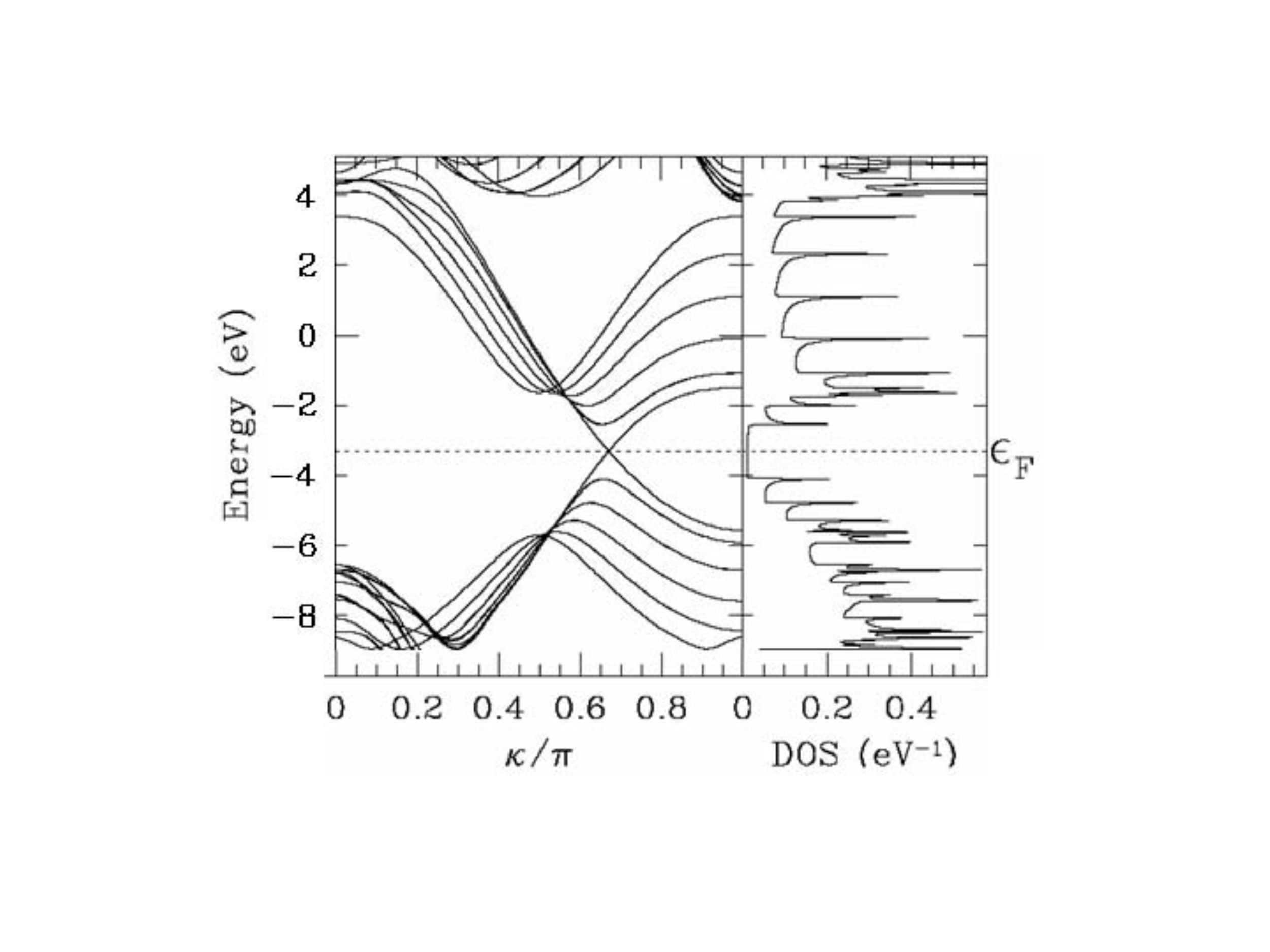}
	\caption{(a) Schematic of the electronic structure of (10,10) armchair nanotubes. (b) Corresponding finite and constant DOS near $E_{F}$. The `gap' associated with DOS peaks at next one-dimensional subbands. From Ref.\ \protect{\cite{Mintmire-3}}}
	\label{armchair}
\end{figure}

From above discussion, the armchair nanotube could be treated as truly 1D metal. The low-energy conduction modes occur near the two gapless Fermi points $\pm\vec{K}=\pm \frac{2\pi}{a}(\frac{2}{3}, 0)$. Excitation of other transversal modes costs the energy $\approx 10$ eV/\textit{n}, and hence a 1D condition arises. For (10, 10) tubes, the conditions for the low-energy regime are met even at room temperature. Up to energy scales $\sim1$ eV, the dispersion relation around the Fermi points remains nearly linear (Fig.\ \ref{armchair} a) and cause a finite and constant density of states (DOS) near $E_{F}$(Fig.\ \ref{armchair} b).

Based on studies of the electronic structures of SWNTs, theoretists also carried out some elementary studies on electronic strutures of MWNTs \cite{Charlier,RSaito3-Chp1,Lambin}. It was argued that a two-layer MWNT constitued by two concentric metallic (or quasimetallic) SWNTs is still metallic (or quasimetallic). An interesting result is, when one of the metallic nanotube is replaced by a semiconducting nanotube, a concentric metal-semiconductor device could be constitute \cite{Benedict}. But there also exists result that indicating changes of electronic properties due to inter-layer interactions when two SWNTs joined together.

Another important research area is on the Mechanical properties of carbon nanotubes. It is well-known that the in-plane C--C covalent bond in graphite is one of the strongest bonds in nature. The best estimate of the in-plane elastic modulus is 1.06 TPa and that of the tensile stiffness is 0.8 TPa. The mechanical properties of ideal carbon nanotubes with negligible number of structural defects should approach or even surpass the reference values of graphite. Moreover, the rolled-up structure of carbon nanotubes would make them withstand repeated bending, buckling, twisting and compression at high rates with complete elasticity.

The hollow structure of carbon nanotubes make it possible to fill them with other materials to form one-dimensional nano-wires. Theoretical studies point out that only materials with small enough surface tensity could be filled into carbon nanotubes, including all the gases and liquids, most of the organic materials and some of the metals and their compounds \cite{Pederson,Galpern,Breton}.

\section{Physical properties, experimental}

Although numerious theoretical calculations have predicted many novel physical and chemical properties for carbon nanotubes, the nanometer size and random orientations of nanotube samples make it extremely difficult to examine and certify these properties experimentally. However, taking advantage of the rapid progress of nano-fabrication and nano-manipulation, scientists are making fast progress on experimental studies and many valuable results has been obtained which agree quite well with theoretical predictions.

Since nanotubes are ideal model systems for the investigation of low-dimensional molecular conductors, measuring the electronic properties of individual nanotubes is always the focus of experimental studies. This is very challenging for two reasons. Both high-quality nanotube samples and new techniques for making electrical contacts to individual tubes are necessary. The first to be studied was individual multiwall nanotubes. By using STM lithography technique, Langer et al. reported the first measurement on individual MWNTs \cite{Langer}. They found that transport properties of MWNTs were consistent with the quantum transport behaviors of weakly disordered low-dimensional conductors. Ebbesen et al. systematically measured four-probe conductance of individual MWNTs and observed both metallic and non-metallic behaviors \cite{Ebbesen1}. Typical temperature dependence of the resistance of metallic tubes was similar to that of disordered semi-metallic graphite, with a moderate negative temperature coefficient. Another elegant method of electrical contact was used by Frank et al. \cite{Frank-Chp1}. A single MWNT attached to a STM tip was repeatly immersed and pulled out of a liquid metal like mercury. Astonishingly, an almost universal quantized conductance at room temperature was measured, providing evidence that transport in MWNTs is ballistic over distances of the order of $\geq1$ $\mu$m.  Recently, a pronounced Aharonov-Bohm resistance oscillation was observed by Bachtold et al. \cite{Bachtold-Chp1}. Their results clearly demonstrate that at low temperatures quantum interference phenomena dominate the magnetoresistance of MWNTs.

Singlewall nanotubes have well defined structures and relatively less defects. The success of synthesis of high-quality SWNTs with uniformed structures greatly stimulated experimental studies. The first results on individual SWNTs were obtained by Tans et al. \cite{Tans-QW1}. At low temperatures step-like current-voltage characteristics were observed that indicate single-electron transport with Coulomb blockade and resonant tunneling through single molecular orbitals. Quite remarkably, the nanotubes appear to behave as coherent quantum wires. The density of states of the molecule consist of well-separated discrete electron states corresponding to an one-dimensional `particle-in-a-box' situation induced by micrometer tube length. The measurement on individual SWNT ropes by Bockrath et al. \cite{Bockrath-QW1} obtained similar results. Soon later, two research groups at Delft Univ. and Harvard Univ. obtained the first atomic resolution STM images of SWNTs and observed the chiral winding of the hexagons along the tube \cite{Wildoer-Chp1,Odom}. The STM was also used to measure the tunneling spectroscopy of nanotubes. They found both metallic and semiconducting spectra. Their data provided the first experimental verification of the bandstructure predictions. The observed gap structures quantitatively agree with the calculations.

The explanations of all these experiments were based on single-electron picture, wherein the Landau Fermi liquid theorem is appropriate. However, the low-dimensional nature of carbon nanotubes would probably introduce pronounced electron-electron interactions, and a non-Fermi liquid behavior such as a Luttinger Liquid is predicted \cite{Egger1-Chp1,Egger2-Chp1,Kane2,Yoshioka-Chp1}. The first generation devices used in transport measurement of individual SWNTs have very high contact resistance, therefore the effect of electron-electron correlations was masked by charging effects. Tans et al. exhibited the first evidence of electron correlations by finding that electrons tunneling into a SWNT are spin polarized \cite{Tans-EE1}. By improving the quality of electric contact, Bockrath et al. recently observed a power-law suppression of tunneling DOS on individual SWNT ropes \cite{Bockrath-LL1}, which could be well explained by tunneling into the bulk and ends of a clear Luttinger Liquid.

The experimental characterizations of mechanical properties of carbon nanotubes is very difficult because of the short length (1--100 $\mu$m) of the samples. Up to now, there are no direct tensile tests available to determine the axial strength of nanotubes. By measuring the amplitudes of intrinsic thermal vibrations of nanotubes during imaging inside a Transmission Electron Microscope (TEM), Treacy et al. estimated a Young's modulus as high as 1.8 TPa which is higher than the in-plane modulus for single-crystal graphite \cite{Treacy}. Recently, Pan et al. reported the first direct tensile test on MWNTs \cite{Pan-tensile}. In their experiment, a tiny stress-strain puller was used to pull a MWNT bundle containing several ten thousand millimeter-long aligned nanotubes, and the average Young's modulus and tensile strength obtained were $0.45\pm 0.23$ TPa and $1.72\pm 0.64$ GPa, respectively. The relatively lower Young's modulus was attributed to the existance of large amounts of defects in CVD derived nanotubes, but it is still higher than that of Steel. Moreover, TEM observations have shown that carbon nanotubes could sustain extreme strain without showing signs of brittleness, plastic deformation or atomic rearrangements and bond rupture \cite{Falvo,Ebbesen2}. The flexibility is related to the ability of the carbon atoms to rehybridize. All these results indicate that nanotubes do have exceptional mechanical properties.

\section{Motivation and goal}

Despite the fast progress in the researches of carbon nanotubes, many important physical properties, such as thermal properties, have not been well experimentally characterized. Success of the synthesis of millimeter-long aligned multiwall carbon nanotubes opened the door of macroscopic manipulations on these nano-sized materials \cite{Pan-Ch1}. The continusity and good alignment of nanotubes enable us to perform measurements on well-defined samples. This dissertation aims at a systematic investigation on MWNTs, including their thermal properties, tunneling spectroscopy and electrical transport properties. 

On thermal properties, our goal is to explore the dimensionality of the phonon structure and its relationship with tube diameter. A potential rewarding experiment would be to measure the specific heat and thermal conductivity of a small sample simultaneously. Fortunately, a so-called $3\omega$ method, which uses a narrow-band technique in detection, and therefore gives a relatively better signal-to-noise ratio, is especially suitable for such measurements. To do this, an explicit solution for the 1D heat-conduction equation of a rod- or filament-like specimen is needed.

On tunneling spectroscopy, remarkably results have been obtained on individual SWNTs, such as chirality related gap structure, one-electron tunneling, Luttinger liquid behavior, etc. However, no systematic measurement has been performed on MWNTs by now. MWNTs have diameters of several tens of nm. This is one order of magnitude higher than that of SWNTs and relates into an order of magnitude difference in energy scale. In zone-folding picture, MWNTs may still be treated as 1D conductor, then an interesting and important problem is whether or not electron correlations exist in MWNTs. Taking advantage of the very long aligned MWNTs, we can use true four-probe configuration to measure the tunneling spectroscopy on a thin bundle sample containing only several tens of nanotubes. Important information will be obtained on the possible effects of electron correlations and their relations with tube diameter.

Finally, electrical transport properties such as thermoelectric power, magnetoresistance will be investigated. For thermoelectric power measurement, the continusity of the sample guarantees the reliability of the results. For magnetoresistance measurement, the larger transverse size of MWNTs compared with SWNTs make them more suitable to explore the orbital effects. A magnetic field of 3T is enough to induce a magnetic field flux of $h/2e$ in a MWNT with outer diameter 30nm. It will be interesting to perform magnetotransport experiments of MWNTs in parallel and perpendicular field at low temperatures to explore quantum transport phenomena.

\putbib[ch1-ref]
\end{bibunit}


\chapter{Sample growth and characterization}

\begin{bibunit}[unsrt]

\setcounter{section}{0}

\section{Introduction}
For the first time, millimeter-long aligned multiwall carbon nanotubes were synthesized using thermal decomposition method by Dr. Z. W. Pan in Prof. S. S. Xie's research group \cite{Pan}. Here a brief introduction of the sample growth and characterization is presented to let the readers have a first impression on the structure and morphology of MWNTs studied in this dissertation. All the materials in this chapter were kindly provided by Prof. S. S. Xie.

\section{Process of synthesis}

The method used is similar to that reported by W. Z. Li et al. \cite{Li}, but some improvement has been made in substrate preparation. Instead of using bulk mesoporous silica containing iron nanoparticles embedded in the pores as substrate, silica film on the surface of which iron/silica nanocomposite particles are evenly positioned was utilized. The iron/silica substrate was prepared by a sol-gel process using the technique described in Ref. [1]. Tetraethoxysilane (10 ml) was mixed with 1.5 M iron nitrate aqueous solution (15 ml) and ethanol (10 ml) by magnetic stirring for 20 min. A few drops of concentrated hydrogen fluoride (0.4 ml) was then added, and the mixture was stirred for another 20 min. The mixture was then dropped onto a quartz plate to form a film of thickness 30--50 $\mu$m. After gelation of the mixture, the gel was dried overnight at $80^{\circ}$C to remove the excess water and other solvent, during which the gel cracked into small pieces of substrates of area 5--20 mm$^2$.

\begin{figure}
	\centering
		\includegraphics[width=0.50\textwidth]{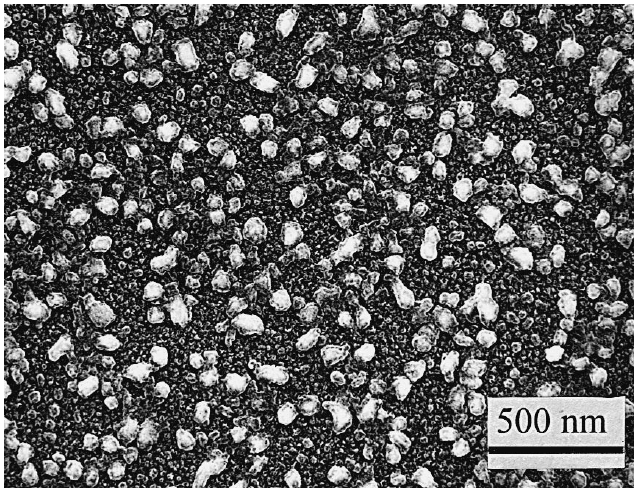}
  \caption{SEM image of the iron/silica substrate after reduction, showing iron/silica nanocomposite particles with size of 5--50 nm present evenly on every surface of the substrate.}
\label{substrate} 
\end{figure}

\begin{figure}
	\centering
		\includegraphics[width=0.40\textwidth,angle=90]{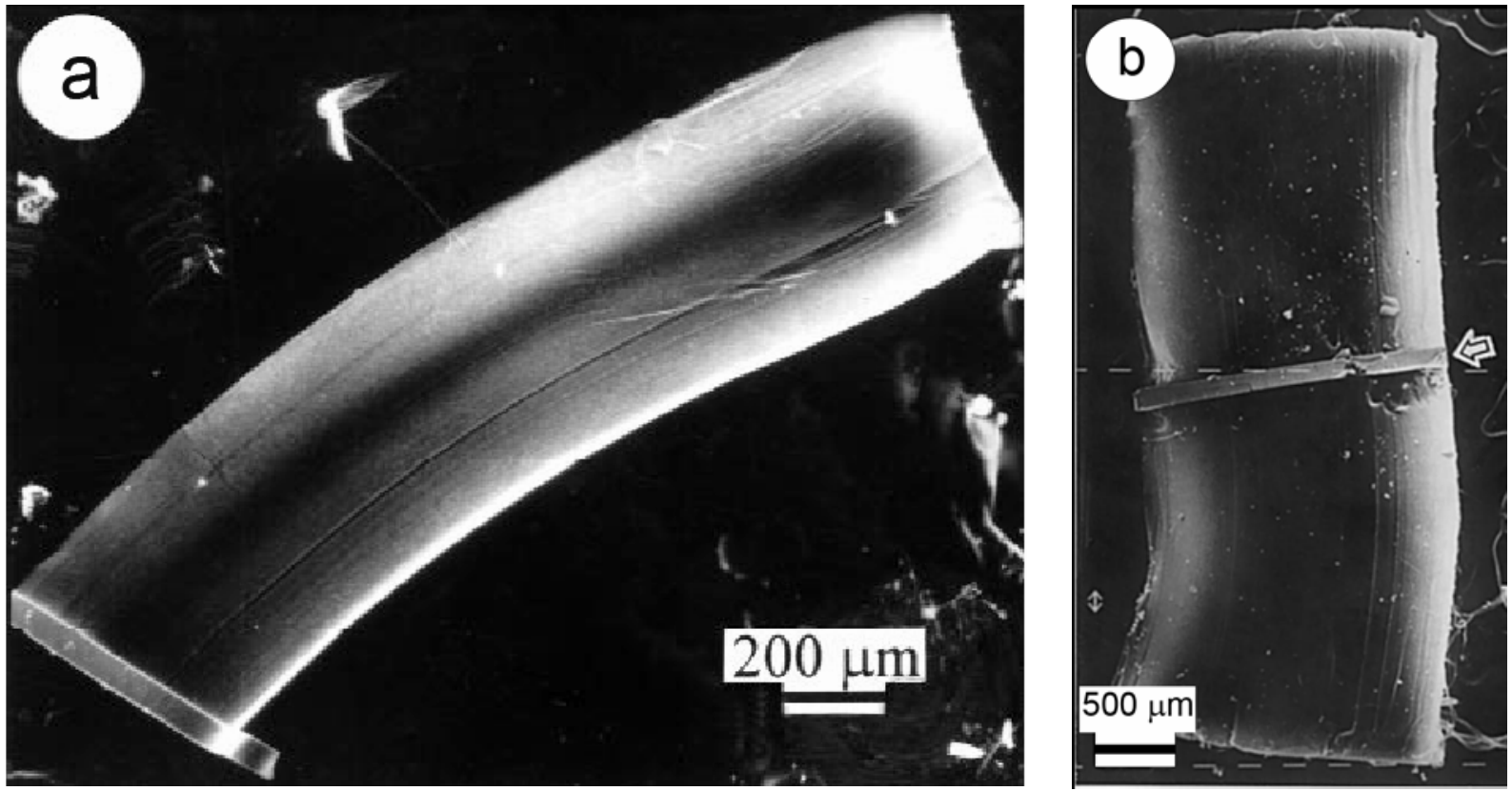}
	\caption{Low-magnification SEM images of a sample after 48 hours of growth. (a) Only one nanotube array remains after stripping off other arrays from the other sides of the substrate. (b) Two nanotube arrays grown from the upper and lower sides of the substrate. From Ref.\ \protect{\cite{Pan}}}
	\label{lowSEM}
\end{figure}

The substrates were placed in a quartz boat and were then introduced into a chamber of a tube furnace. The substrate were calcined at $450^{\circ}$C for 10 h under vacuum ($<10^{-3}$ Torr) and then reduced at $500^{\circ}$C for 5 h in a flow of 9\% hydrogen in nitrogen under 180 Torr. At this stage, large quantities of nanoparticles with size of 5--50 nm formed evenly on all surfaces of the substrates (Fig.\ \ref{substrate}). Energy-dispersive X-ray (EDX) spectra taken from these nanoparticles showed the presence of iron, silicon and oxygen (68.7, 14.5 and 16.8 wt\%, respectively), which indicated that these nanoparticles were iron/silica nanocomposite particles. The iron/silica particles were believed to act as catalysts for nanotube growth. Subsequently, a flow of 9\% acetylene in nitrogen was introduced into the chamber at a flow rate of 110 cm$^{3}$/min, and carbon nanotubes were formed on the substrate by deposition of carbon atoms from decomposition of acetylene at $600^{\circ}$C under 180 Torr (Fig.\ \ref{lowSEM}). The growth time varied from 1 to 48 h.

The as-grown nanotubes were examined by a scanning electron microscope (SEM; S-4200, Hitachi), and EDX spectra were recorded by a SiLi detector attached to the SEM. A transmission electron microscope (TEM; JEOL JEM-200cx at 200 keV) was used to characterize the fine structures of nanotubes.

\section{Characterization}

\subsection*{SEM and TEM}

\begin{figure}
	\centering
		\includegraphics[width=0.90\textwidth,angle=90]{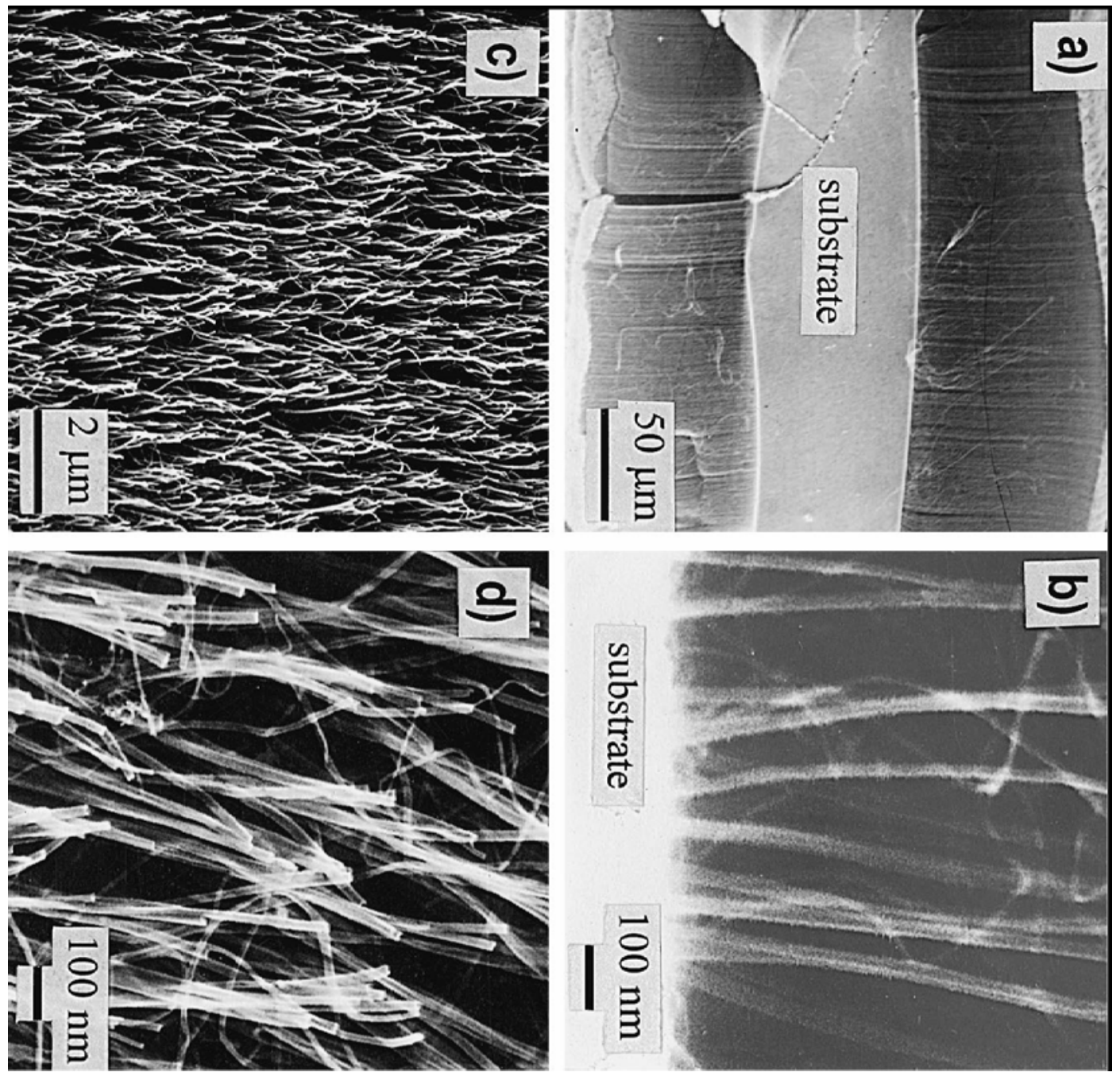}
	\caption{High-magnification SEM images of aligned carbon nanotube arrays. (a) Aligned nanotubes grow outwards perpendicularly from the surface of the substrate to form nanotube arrays. (b) Higher magnification of the nanotube array showing aligned nanotubes have uniform external diameters (20--40 nm) and spacing ($\sim 100$ nm) between tubes. (c, d) Bottom structure of the nanotube array showing well-separated, aligned and open nanotubes. From Ref.\ \protect{\cite{Pan2}}}
	\label{highSEM}
\end{figure}

High-magnification SEM image (Fig.\ \ref{highSEM}a) shows that carbon nanotubes grow out separately and perpendicularly from the substrate to form an array. The nanotubes within the array are of uniform external diameter (20--40 nm) and spacing ($\sim$ 100 nm) between tubes (Fig.\ \ref{highSEM}b). Most of the nanotubes in the array are highly aligned, although a few of them appear to be slightly tangled or curved. We note that no traces of polyhedral particles or other graphitic nanostructures are detected in both the bottom part and central part of the array, which indicates that the nanotubes prepared in this study have very high purity and thus have very high quality.

The nanotube array can be easily stripped off from the substrate without destroying the array's integrity, and the SEM images taken from the bottom end of the array confirm that the nanotubes are highly aligned and well separated (Fig.\ \ref{highSEM}c, d). EDX spectra collected from the bottom end of the array demonstrate the presence of carbon alone, neither silicon nor iron could be detected. 

\begin{figure}
	\centering
		\includegraphics[width=0.50\textwidth]{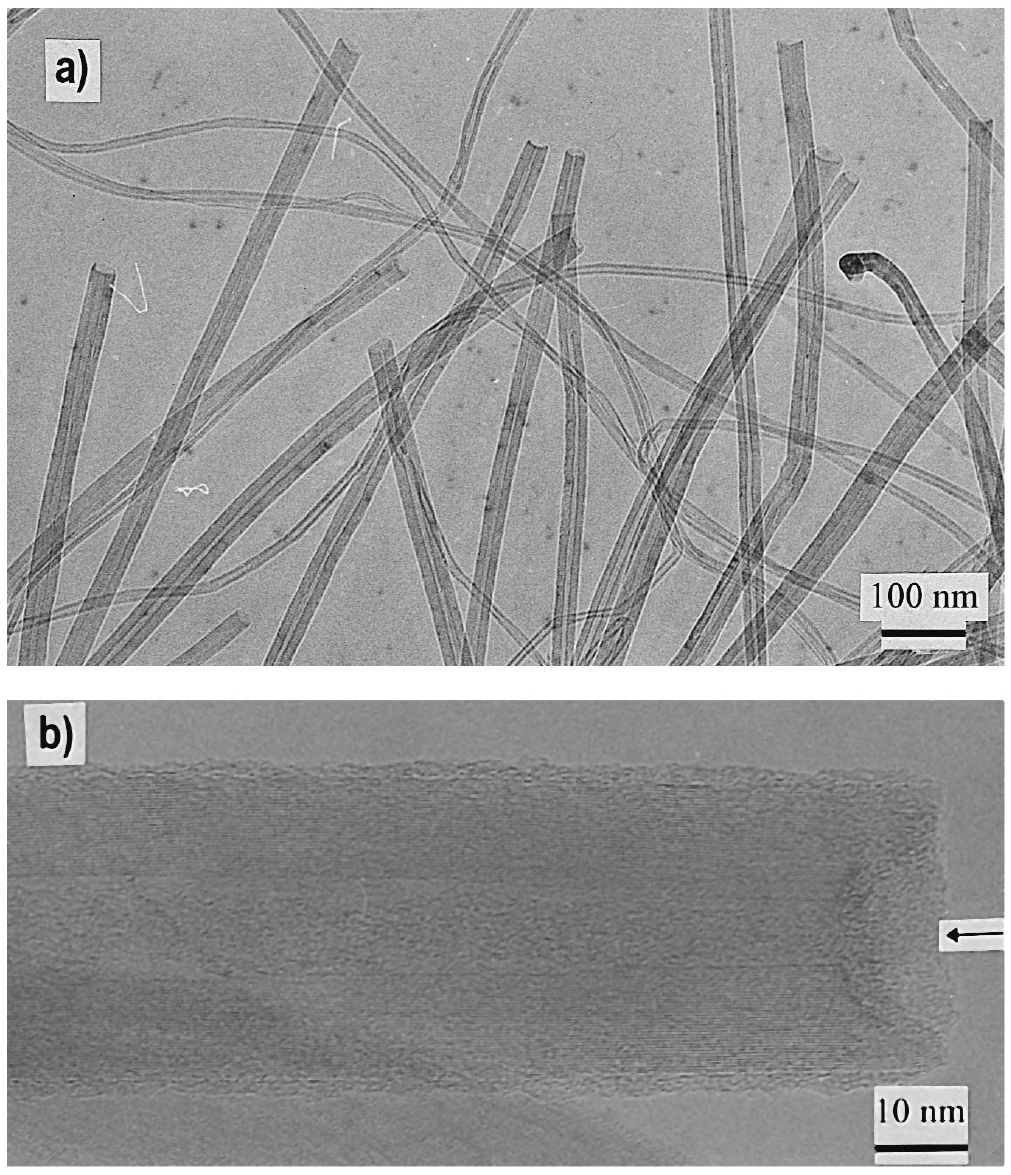}
	\caption{(a) Low-resolution TEM image of the bottom end of the nanotube array showing aligned open carbon nanotubes. Encapsulated iron/silica nanocomposite particles are absent. (b) High-resolution TEM image of the bottom end of an open nanotube showing bowl-shape
structure (see arrow) formed by the graphite sheets. From Ref.\ \protect{\cite{Pan2}}}
	\label{TEM}
\end{figure}

Low-resolution TEM observations of the bottom end of the nanotube array reveal that no seamless caps exist at the bottom ends of the nanotubes and the tubes are indeed opened (Fig.\ \ref{TEM}). The inner and outer diameters of the open nanotubes are 10--15 and 20--40 nm, respectively. Encapsulated particles are not observed at the bottom ends of the nanotubes. High-resolution TEM observations confirm that the nanotubes are opened at the bottom ends. The nanotubes are well graphitized and typically consist of 10--30 concentric graphite layers. 

\subsection*{Raman spectroscopy}

Raman spectroscopy is a very useful tool to characterize carbon materials because it gives information on the hybridization ($sp^n$) state, the size of the graphite crystallites and the degree of ordering of the material. Crystalline graphite, such as highly oriented pyrolitic graphite (HOPG), belongs to the space group $D_{6h}^{4}$ with the vibrational modes having the irreducible representation $\Gamma=2E_{2g}+E_{u}+A_{2u}+2B_{2g}$. Only the two $E_{2g}$ modes are Raman active. These correspond to the movement of two neighbouring carbon atoms in a graphene sheet. HOPG is characterized by a single strong band at 1580 cm$^{-1}$ which is one of these $E_{2g}$ modes. On the other hand, glassy carbon has its most intense band at approximately 1350 cm$^{-1}$ which is normally explained by the relaxation of the wave vector selection rule due to the finite size of the crystallites in the material.

For raw MWNTs made by the arc-discharge method, the first-order Raman spectra has a weak and broad peak at 1346 cm$^{-1}$ (D band) and a strong peak at 1574 cm$^{-1}$ (G band) \cite{Hiura}. The G band is very similar to 1580 cm$^{-1}$ G band of HOPG but is slightly downshifted perhaps due to the curvature in the carbon network. The width of this peak is almost same to that of HOPG (approximately 23 cm$^{-1}$), suggesting a highly crystalline graphitic structure. The 1346 cm$^{-1}$ peak is a disorder-induced peak in graphitic structures and in nanotubes it could arise from the smaller crystallite sizes, defects in the curved graphene sheets or at the tube ends. Considering the Raman features, MWNTs must be seen as graphite-like microcrystals and are unlike fullerenes which show vibrational features of molecules.

Z. W. Pan et al. studied micro-Raman spectroscopy of arrays of high-density aligned MWNTs made by thermal decomposition method and compared their results with that of HOPG and MWNTs made by arc-discharge method \cite{Pan3}. Here is a brief description of their results.

Raman scattering experiments were performed at ambient conditions by a micro-Raman spectrometer (Reinshaw, England) in a backscattering geometry. The spectra were collected by using 514.5 nm excitation of argon laser at 0.2 mW and a spectral slitwidth $\leq 2$ cm$^{-1}$, with the laser incident perpendicular to the orientation of nanotube arrays. The diameter of focus area of laser beam is about 2 $\mu$m. The time for collecting every spectrum is 5s $\times$ 3.

\begin{figure}
	\centering
		\includegraphics[width=0.70\textwidth]{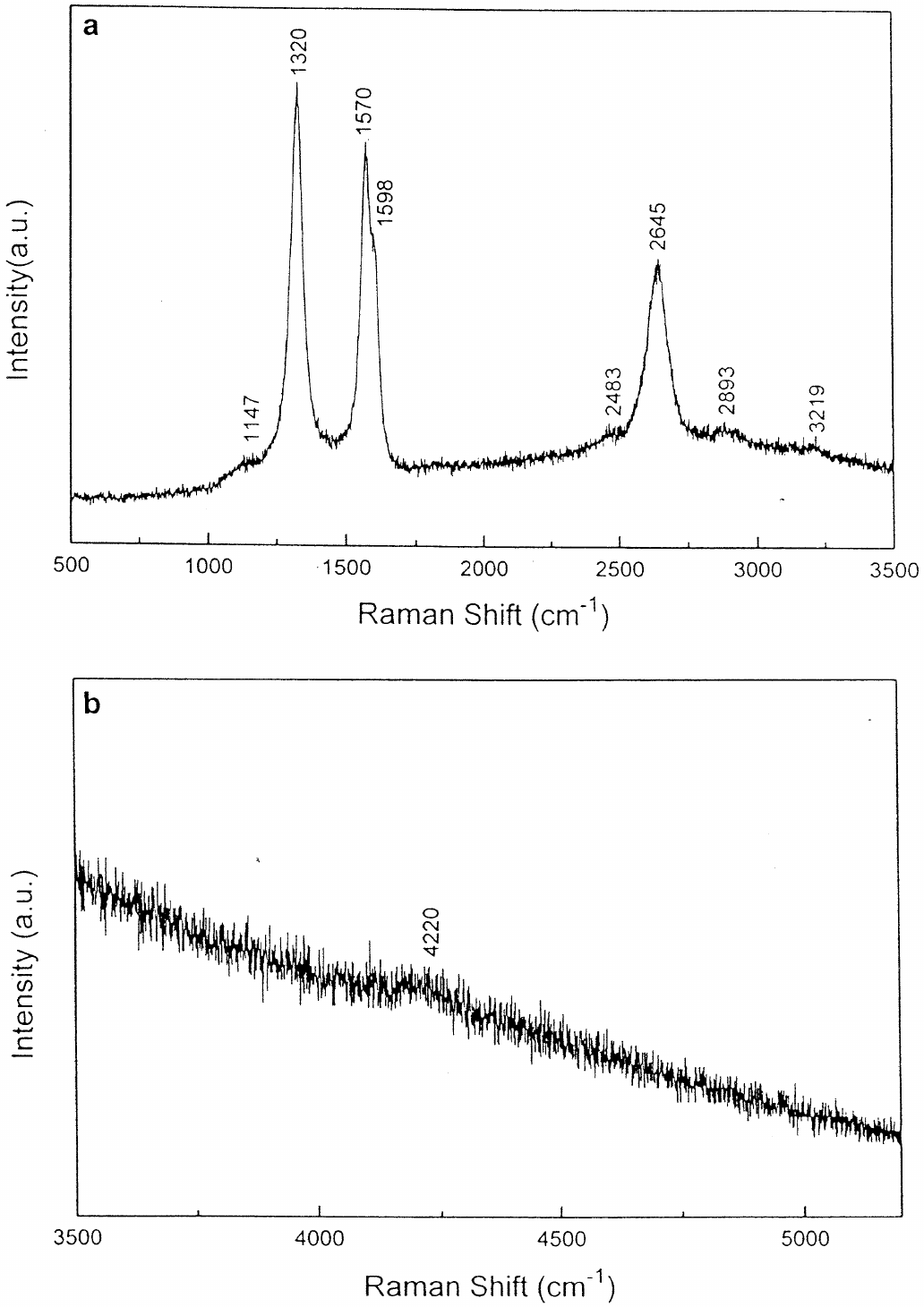}
	\caption{Micro-Raman spectra of aligned multiwall carbon nanotubes. (a) first-order and second-order Raman spectra. (b) Third-order Raman spectrum. From Ref.\ \protect{\cite{Pan3}}}
	\label{raman}
\end{figure}

Shown in Fig.\ \ref{raman} are Raman spectra of aligned MWNTs. Fig.\ \ref{raman}a are first-order and second-order Raman spectra, Fig.\ \ref{raman}b is third-order Raman spectrum. No smoothing was applied to the spectra.
In Fig.\ \ref{raman}a, the 1570 cm$^{-1}$ peak (G band) is the high-frequency Raman active $E_{2g}$ mode. It is downshifted by 11 cm$^{-1}$ compared with G band (1581 cm$^{-1}$) of HOPG, but is close to G band (1574 cm$^{-1}$) of MWNTs made by arc-discharge method. Remarkablely, a strong peak can be seen at 1320 cm$^{-1}$ (D band), and a weak peak at 1598 cm$^{-1}$ (D$'$ band) appears. It is noticeable that the strong 1320 cm$^{-1}$ peak resembles the D band of pyrolitic graphite (PG), but can not be observed in HOPG. In other carbon materials, the appearance of D and D$'$ band is attributed to the crystallite-size effect or lattice twisting. The ratio of intensity of D band relative to G band reflects the size of graphitized crystallites. In SEM and TEM studies, no carbon nano-particles was found. Therefore, the appearance of 1320 cm$^{-1}$ peak could be attributed to defects in the curved graphene sheets or at the tube ends, turbostratic structure in graphene sheets, limited size of crystal domains in nanotubes and the coating of amorphous carbon at nanotube surfaces. Compared with MWNTs made by arc-discharge method, the intensity of D band is much higher. This mainly due to the much lower growth temperature of pyrolysis method (600$^{\circ}$C). The results of Raman spectroscopy show that MWNTs made by pyrolysis method have relatively lower degrees of graphitization and larger amounts of defects.

In the second-order Raman spectrum, four peaks at 2645 cm$^{-1}$, 2893 cm$^{-1}$, 2483 cm$^{-1}$ and 3219 cm$^{-1}$ can bee seen. The 2645 cm$^{-1}$ band is the second-order spectrum of D band ($\sim2\times1320$ cm$^{-1}$), the 2893 cm$^{-1}$ band is identified as composition of D band and G band (1320 cm$^{-1}+1570$ cm$^{-1}$).

In the third-order Raman spectrum, a weak 4220 cm$^{-1}$ peak can be recognized. This peak is the third-order Raman spectrum of nanotubes and can be identified as composition of D band and G band ($2\times1320$ cm$^{-1}+1570$ cm$^{-1}$). It is downshifted compared with 4320 cm$^{-1}$ peak of HOPG and 4305 cm$^{-1}$ peak of PG. No fourth-order Raman spectrum was observed in this experiment.

From the above analysis, the Raman feature of MWNTs prepared has a strong resemblance to that of PG. This indicates that many nanometer-sized crystal domains exist in microscopic structure of MWNTs. The appearance of the third-order Raman spectrum shows that these crystal domains have high degrees of crystallization, which agrees with the conclusion from TEM studies.

\section{Summary}

Aligned and open-ended multiwall carbon nanotubes (MWNTs) were prepared by Dr. Z. W. Pan at a very high yield by pyrolysis of acetylene over film-like iron/silica substrates. These nanotubes grow outwards separately and perpendicularly from the surfaces of the substrates to form aligned nanotube arrays. The nanotubes within the arrays are of uniform outer diameters (20--40 nm), with a spacing of $\sim100$ nm between the tubes. After continuous growth of 48 hours, the lengths of nanotubes can reach 2--3 mm, which is 1--2 orders higher than previously reported values. TEM studies indicate a high degree of graphitization. The multiwall nanotubes are composed of 5--30 concentric graphene cylinders with average inter-wall distance of 0.34 nm. The inner diameters of nanotubes are 3--10 nm. Features of micro-Raman spectra show a strong resemblance to that of Pyrolytic Graphite. The appearance of a strong D band indicates the existance of nanometer-sized crystal domains in MWNTs. The appearance of the third-order Raman spectrum shows that MWNTs have high degrees of crystallization.

\putbib[ch2-ref]
\end{bibunit}


\chapter{Thermal properties}

\begin{bibunit}[unsrt]

\setcounter{section}{0}

\section{Introduction}

While the electronic properties of carbon nanotubes have  
been intensively studied and shown to be peculiar,
such as the sensitive dependence of the  
electronic structure on the diameter and chirality of the  
tubule, the thermal properties of the material are not well  
characterized experimentally, though they are as important  
as the former both from the viewpoint of basic research and for possible  
application. Benedict {\it et al.} \cite{Benedict-Chp3} predicted  
that the temperature dependence of specific heat would  
depend on the diameter of the tubule, and might even show  
a sign of dimensional cross-over as the temperature is varied.  
However, an experimental test of the thermal properties of the  
material is frustrated because the samples usually available  
are in a mat-like form, in which each individual tubule, or  
rope of tubes, is rather short and only loosely connected  
to each other, which prevents a reliable and  
equilibrium thermal measurement. The recent success in  
synthesis of millimeter-long highly aligned array of MWNTs  
\cite{Pan1} enables us to perform specific heat and  
thermal conductivity measurements on well-defined samples.

\section{Experimental} 
The MWNTs were synthesized through chemical vapor  
deposition (CVD) \cite{Pan1}. High magnification scanning electron  
microscopy (SEM) analysis shows that the tubules grow out  
perpendicularly from the substrate and are evenly spaced at an  
averaged inter-tubule distance of $\sim$100 nm, forming a  
highly aligned array. High resolution transmission electron  
microscopy (HRTEM) study shows that most of the tubules are within 
a diameter range of 20--40 nm. The mean external diameter 
is $\sim$30 nm. A tubule may contain 10--30  
walls, depending on its external diameter. 
The continuity of the tubules, which is of crucial importance for  
the interpretation of the thermal conductivity data, is  
indicated by SEM studies, and by Young's modulus  
measurement where several tenth of TPa, a value higher than  
that of steel, was reached \cite{Pan2-Chp3}. The final samples  
used in our measurements were 1--2 mm long bundles of  
MWNTs stripped off from the bulk array. The apparent  
cross-section of the samples spreads from 10$^{-10}$ to  
10$^{-8}$ m$^2$. The filling factor of MWNTs in the bundles  
is $\sim$1.5\%, estimated from SEM observation. This  
estimation may bring some uncertainty in determining the  
absolute values of the thermal conductivity and specific heat  
of the MWNTs. However, their temperature dependence is  
not affected by this uncertainty. 

\begin{figure}
	\centering
		\includegraphics[width=0.70\textwidth]{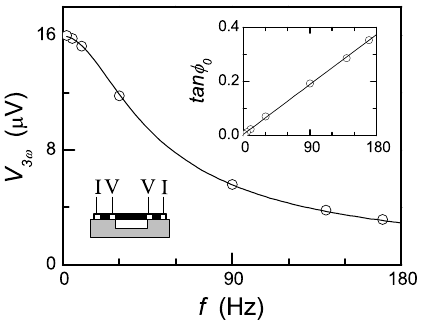}
	\caption{The frequency dependence of the  
amplitude (main frame) and the phase shift $\phi_0$ (upper  
right inset) of the 3$\omega$ voltage measured on one of  
the samples at $T=50$ K. Solid lines are the predictions of  
Eqs. 1 and 2. Lower left inset: schematic diagram of sample  
mounting. The samples were glued to sapphire substrate  
with silver paint in such a way that the inner part between  
two voltage contacts was suspended. The silver paint  
contacts serve both as electrodes and heat sinkers of the  
sample. The $3\omega$ voltage was measured by using a  
lock-in amplifier (SR850). The whole set-up was 
heat-shielded to the substrate temperature and was maintained in  
high vacuum, to preclude heat leakage through air. The  
total heating power at each measured temperature was  
adjusted to be small comparing to the value of sample's  
thermal conductance. The heat leakage through radiation is  
estimated to be less than 5\%.}
\label{setup} 
\end{figure}

Measurements of  
thermal conductivity and specific heat on such tiny but well  
defined samples were made possible by using a self-heating  
$3\omega$ method. The basic idea of the method can be  
traced back to 1910 \cite{Corbino}, and was carefully  
tested later on. If both voltage contacts of the sample are  
ideally heat sunk to the substrate, but keeping the 
in-between part suspended (illustrated in Fig.\ \ref{setup}), an {\it ac}  
current of the form $I_0sin\omega t$ passing through the  
sample will create a temperature fluctuation at $2\omega$,  
which will further cause $3\omega$ voltage harmonics,  
$V_{3\omega}$, across the voltage contacts.  
$V_{3\omega}$ can be solved explicitly to an accuracy of  
$\sim$1/81 \cite{Lu}: 
\begin{equation} 
V_{3\omega}=\frac {2I_0^3LR\left( dR/dT\right)} {\pi^4\kappa  
S\sqrt{1+(2\omega \gamma)^2}} sin(3\omega t-\phi_0)  
\label{1} 
\end{equation} 
\begin{equation}
tan\phi_0=2\omega \gamma 
\label{2} 
\end{equation} 
where $\gamma=L^2/\pi^2a^2$, $a^2=\kappa/C_p\rho_m$ is the  
diffusivity coefficient, $\kappa$, the thermal conductivity,  
$C_p$, the specific heat, $\rho_m$, the density, $R$, the  
resistance, $L$, the length (between voltage contacts), and  
$S$, the cross-section of the sample. By measuring the  
frequency dependencies of the amplitude and the phase shift  
$\phi_0$ of $V_{3\omega}$ in proper $\omega$ and $I_0$  
ranges (typical results are shown in Fig.\ \ref{setup}), both $\kappa$,  
$a^2$, and hence $C_p$ of the sample can be determined.  
This method was verified to be reliable on pure Pt-wire  
samples. The measured $C_p$ of Pt agrees with the  
standard data within an accuracy of 5\% from 10 K to 300  
K.

\section{Results and discussion}

\begin{figure}
	\centering
		\includegraphics[width=1.00\textwidth]{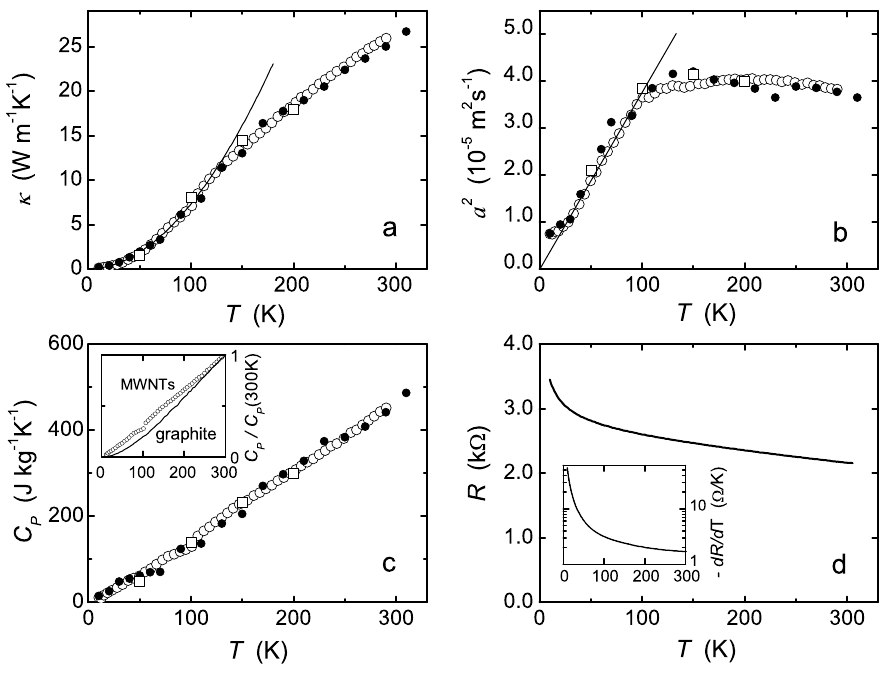}
    \caption{(a) Temperature  
dependence of the thermal conductivity of three MWNT  
samples. It is roughly linear above $\sim$120 K, but  
becomes quadratic at temperatures below: $\kappa\propto 
T^{1.98\pm 0.03}$ (the solid line). (b)  
The thermal diffusivity coefficient of the samples. (c)  
The specific heat of the samples, calculated using the data in  
figures (a) and (b), showing a linear temperature  
dependence, which differs significantly from that of  
graphite \protect{\cite{DeSorbo}} (see the inset).
For graphite, $C_p$(300K) $\simeq720$ Wm$^{-1}$K$^{-1}$. (d) 
Typical resistance $R$ (main frame) and $dR/dT$ (inset) of the samples, 
which are used by Eq. 3.1 to calculate  
the $\kappa$ and $a^2$.} 
\label{results} 
\end{figure}

Shown in Fig.\ \ref{results} are the main results. First we notice that,  
while both $\kappa$ and $a^2$ are significantly non-linear,  
$C_p$ calculated from them follows a pretty linear 
$T$-dependence over the entire temperature range measured.  
We note that a $T$-linear  
phonon specific heat was predicted for some special  
systems, but seldom observed experimentally over such a wide  
temperature range. 
This behavior is dominated by the phonon contribution, since  
the electron contribution is negligible in the temperature  
range of this experiment, as expected from the fact that the  
electron density of states is either very low or gapped at the  
Fermi surface of carbon nanotubes \cite{Wallace,Mintmire2,Kane-Chp3}.

The phonon contribution can generally be written as:  
\begin{equation}  
C_p=\int_{0}^{\omega_{max}}
{k_B
\left(\frac {\hbar\omega} {k_BT}\right)^2 
\frac {e^{\frac {\hbar\omega} {k_BT}}\rho(\omega)d\omega} 
{( e^{\frac {\hbar\omega} {k_BT}}-1 )^2}}
\label{3}  
\end{equation} 
where $\rho(\omega)$ is the phonon  
spectrum determined by the phonon energy dispersion  
$\omega(k)$ of different modes, and by the occupational  
dimensionality of phonon excitations in $k$-space.  
Detailed explanation of the $T$-dependence of $C_p$  
requires the information on the $\rho(\omega)$ of the MWNTs.  
While the $\rho(\omega)$ and $\omega(k)$ for single-wall  
tubules of $\sim$1 nm diameter have been calculated  
numerically and shown to have a linear phonon dispersion 
and 1D behavior \cite{Jishi,Yu,RSaito3}, such calculation  
is virtually impossible to be performed on tubules of 
a few tens nm diameters, containing over hundreds of  
hexagonal cells along the circumference. Nevertheless, one  
can still find immediately from Eq. 3.3 that a constant  
$\rho(\omega)$ would result in a linear $T$-dependence of  
$C_p$ when $T\ll\Theta_D=\hbar \omega_{max}/k_B$,  
the Debye temperature. This condition is believed to  
hold in our case, referring to the fact that the C--C bond  
configuration within the wall is pretty much the same as in
graphite, hence its $\Theta_D$ should not be far  
from $\sim$2400 K \cite{DeSorbo}.  
Therefore, we conclude that the linear $T$-dependence of  
$C_p$ represents a constant $\rho(\omega)$ 
of the MWNTs for the phonon states excitable in the 
temperature range of this experiment. It is less likely that a complicated  
$\rho(\omega)$ could ``happen" to result in a linear  
$C_p$ after the integration, in this case one would usually  
get a complicated $T$-dependence of $C_p$. 
Strictly speaking, due to the  
rolled-up nature of the material, there should exist some 
spiky-like structures over the flat background of MWNTs'  
$\rho(\omega)$. Such structure is most significantly seen in  
tubules with smallest diameter \cite{Yu}. Nevertheless, thermal  
properties like $C_p$ are integration over a wide frequency range  
and hence would be rather insensitive to the local spiky  
structures.  
 
A constant $\rho(\omega)$ usually originates from linearly
dispersed branches in a 1D system, or from one or more 
quadratically dispersed branches in a 2D system. The dimensionality
of our MWNTs can be determined
through the following criteria \cite{Benedict-Chp3}. Because of  
the rolled up structure, the allowable phonon states are  
quantized in $k$-space, forming some discrete lines  
parallel to the tubule axis. A tubule is  
strictly 1D-like if only those states on the central line are  
thermally excitable, otherwise if many lines are occupied  
with phonons then the tubule should be 2D-like. 
A simple estimation shows that even at the  
lowest temperature of this experiment ($\sim$10 K), and  
even for the inner-most wall of our MWNTs (which has the largest 
inter-line distance), the thermally  
excitable phonon states would cover several to several tens of  
lines, depending on whether a linear or a quadratic dispersion is
assumed. Therefore, our MWNT at higher temperatures 
({\it i.e.}, $\geq$10 K) can approximately be
regarded as a 2D system with a constant $\rho(\omega)$. This 
approximation sounds better if 
considering the fact that the outer walls, which have larger 
weight on the total $C_p$ than the inner walls, undergo 
dimensional cross-over at even lower temperatures.

\begin{figure}
	\centering
		\includegraphics[width=0.70\textwidth]{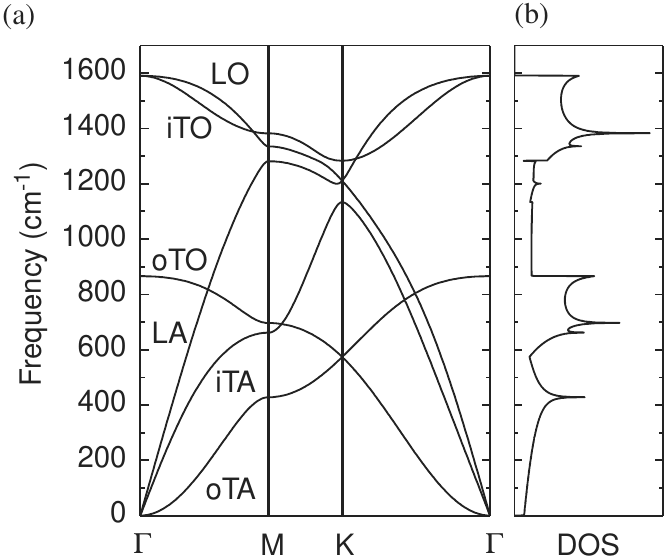}
\caption{Graphite phonon dispersion relations and the corresponding phonon density of states. From Ref.\ \protect{\cite{Dresselhaus05}}}
\label{graphite} 
\end{figure}

Such a phonon structure is similar to, but still different from  
that of graphite whose $\rho(\omega)$ is only roughly flat  
in a moderate frequency range \cite{Nicklow}. The similarity is easy to  
understand, for that graphite is the parent material of  
MWNTs. The central issue now is how to understand the  
difference in $\rho(\omega)$ induced by the rolling up of  
graphene sheets. For flat sheet, as in graphite, 
its hexagonal symmetry  
leads to a quadratically dispersed acoustic phonon branch,  
corresponding to the out-of-plane vibration mode of the  
sheet. This branch contributes to a constant term in  
$\rho(\omega)$ \cite{Nicklow,Komatsu1,Komatsu2,Landau} (Fig.\ \ref{graphite}). 
However, there are at least two reasons  
that prevent graphite from strictly demonstrating a constant  
$\rho(\omega)$ and a linear $C_p$. First, inter-layer  
coupling could not be neglected at low frequencies, which  
gives $\rho(\omega) \sim \omega$ and $C_p \sim T^2$ (in  
the temperature range of interest here we do not consider  
the case of very low frequencies). Second, the other two  
linearly dispersed in-plane branches, which gives  
$\rho(\omega) \sim \omega$ and $C_p \sim T^2$, also contribute 
to the total specific heat \cite{Komatsu1,Komatsu2}. 
Therefore, no strictly $T$-linear $C_p$ as well  
as strictly flat $\rho(\omega)$ exist \cite{DeSorbo}.

From the above discussion, the fact that MWNTs' linear $C_p(T)$ 
extrapolates to the (0, 0) point seems to
indicate that the out-of-plane acoustic mode (as  
in a graphene sheet) dominates the thermal  
properties of this 2D-like material, and that
the two in-plane acoustic modes, presumably linearly dispersed, 
give a negligible contribution. Here an assumption has  
spontaneously been made: for MWNT of a few tens nm diameter 
its acoustic phonon modes 
can still be classified as one LA and two TAs as in graphite.
This assumption should be true, for that at most temperatures of this 
experiment the wavelengths of the majority of phonons  
are much shorter than tubules' diameter, {\it i.e.}, many  
lines in $k$-space are occupied with phonons. 
The striking  
linearity of $C_p$ further indicates that the coupling  
between the walls in MWNTs is much weaker than that  
in graphite, so that one can treat a MWNT
as a few decoupled single-wall tubules, as far as their
vibrational properties are concerned. 
 
Two facts may be responsible for the weak inter-wall  
coupling: the larger inter-wall distance in MWNT than the  
inter-layer distance in graphite, and the turbostratic stacking  
of adjacent walls which is unavoidable in the rolled-up  
structures. Although there are a number of experiments indicating
that the inter-wall distance in MWNT is larger than that in 
graphite \cite{YSaito,Sun}, recent HRTEM study \cite{Kiang} 
shows that the inter-wall distance decreases as 
tubule's diameter increases, at diameters $\geq$10 nm 
it saturates to $\sim$0.344 nm, a characteristic 
inter-layer distance in turbostratic stacking.
Therefore, we believe that the weak  
inter-wall coupling in MWNT is rather caused by the  
turbostratic stacking of adjacent walls. 

The thermal conductivity of our MWNTs is rather  
low compared to what is generally expected \cite{Service}.  
Since the apparent Wiedemann-Franz ratio 
$R(\kappa S/L)/T$ is about two orders of magnitude larger
than the free electron Lorenz number, the measured $\kappa$ 
essentially reflects the phonons 
contribution, similar to the cases in graphite 
and carbon fibers. Whereas $\kappa$ of graphite shows a marked 
maximum around 100 K, $\kappa$ of our MWNTs monotonously  
decreases with lowering temperature. If roughly expressing  
$\kappa$ as $\kappa \sim T^n$, then $n$ is a little less then  
unity above $\sim$120 K, resulting in a faint hump of  
diffusivity $a^2$ between 120--300 K. Below $\sim$120  
K, $\kappa$ follows almost a quadratic $T$-dependence,  
{\it i.e.}, $n \simeq 1.98$, leading to a quick decrease of  
$a^2$, or to say, a nearly linear decrease of apparent  
phonon mean-free-path with temperature. 
This behavior indicates that drastic change in phonon scattering
mechanism occurs at $\sim$120 K.
Comparing with graphite, although the $\kappa$ of the latter also 
roughly takes a quadratic temperature dependence at low temperatures, 
the mean-free-path of phonons in graphite is rather confined
by boundary scattering, because the $C_p$ of graphite 
is quadratically temperature-dependent at low temperatures
({\it i.e.}, below 70 K \cite{Kelly}). 
An energy-dependent mean-free-path in MWNTs contradicts to 
the usual boundary scattering mechanism seen in graphite 
and other solids at low temperatures. 
One difference between nanotubes relative to graphite sheets is the
existence of the radial breathing mode in the former \cite{Rao}. 
Perhaps the thermal excitation of such phonons is responsible for the
change in scattering mechanism around 120 K. Further investigation 
is needed to clarify the issue.

\begin{figure}
	\centering
		\includegraphics[width=0.70\textwidth]{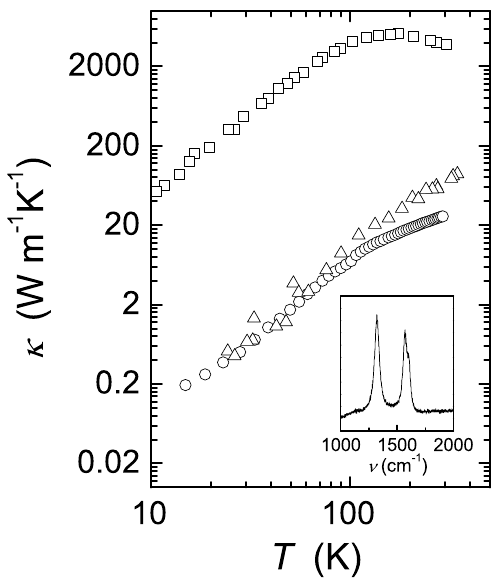}
\caption{A log-log plot of the thermal conductivity of 
our samples (circles) comparing with that of vapor-grown carbon fibers 
\protect{\cite{Heremans}} (triangles: as-grown, 
squares: heat treated at 3000 $^\circ$C).
Inset: First-order Raman spectrum of our MWNTs.}
\label{fiber} 
\end{figure}

The MWNTs grown by CVD method at temperatures as low as $\sim$600 K
are far from perfect, as indicated by the fact that
their thermal conductivity and electrical 
conductivity are about two orders of magnitude lower than those of 
perfect crystalline graphite at room-temperature. The average electrical 
resistivity of MWNTs was estimated by assuming that all walls in 
a MWNT conduct electric current, the characteristic value is about 
$1\times10^{-4}$ $\Omega$m at 300 K. This value, together with the weak 
negative temperature-dependent coefficient, are similar to those of 
disordered semimetallic graphite, and are comparable to 
previously reported data on MWNT bundles \cite{Song,Langer,Dai}.
Figure\ \ref{fiber} shows that the $T$-dependent behavior of  
MWNTs resembles those of not well graphitized materials  
such as the as-grown carbon fibers and glassy carbons 
\cite{Heremans},  
whose in-plane domain size of the hexagonal ordering is  
only a few nm. It has been shown in those materials that the  
mean-free-path of the phonons deduced from the $\kappa$  
data is well correlated with the domain size  
\cite{Heremans,Nysten}. We therefore speculate  
that the domain size of our CVD-derived MWNTs is the  
same as the phonon mean-free-path $l$, which is a few nm  
if estimated using the classical relation $\kappa \sim  
C_pvl$ and assuming a characteristic sound velocity $v \sim  
10^4$ m/s. This picture is further supported by  
micro-Raman investigation wherein the 1320 cm$^{-1}$  
band, which is prohibited in perfect graphite,   
becomes substantially high in our MWNTs, resembling the case 
of as-grown CVD carbon fibers \cite{Heremans}. 
Nevertheless, the gross phonon spectrum is unlikely to 
be significantly altered upon the reduction of the 
domain size to a few nm \cite{Heremans}. We also note that for disordered systems, 
a two-level mechanism could contribute to a linear term of 
specific heat additive to the ordinary term and even 
surpass the latter at very low temperatures \cite{Phillips}.  
Obviously, such a mechanism, if exists, cannot account for a  
total linear $C_p$ of the MWNT over the entire temperature  
range measured. Therefore, our previous discussions on the $C_p$ and the phonon 
structure of the MWNTs should still be valid. 

\section{Summary}

The specific heat and thermal conductivity of millimeter-long 
aligned carbon multiwall nanotubes (MWNTs) have been measured. 
As a rolled-up version of graphene  
sheets, MWNT of a few tens nm diameter is found to  
demonstrate a strikingly linear temperature-dependent  
specific heat over the entire temperature range measured
($10-300$ K). The results indicate that  
inter-wall coupling in MWNT is rather weak compared  
with in its parent form --- graphite, so that one can treat
a MWNT as a few decoupled two-dimensional (2D) single wall 
tubules. The thermal conductivity 
is found to be low, indicating the  
existence of substantial amount of defects in the MWNTs
prepared by chemical-vapor-deposition method.

\vspace{0.5in}

\section{Appendix: $3\Omega$ method}

We present here a novel and simple way to measure the specific heat and thermal conductivity of rod- or filament-like specimens. This method uses four-probe configuration same as for measuring electrical resistance, except for that the voltage contacts serves not only as electrical contacts but also as thermal contacts, to heat-sink the specimen at these two points to the substrate temperature. The specimen, which needs to be electrically conductive and with a temperature-dependent resistance, serves both as heater and temperature sensor to tell the thermal response of the material. With this method we have successfully reproduced the specific heat and thermal conductivity of Pt-wire specimens, and measured the specific heat and thermal conductivity of tiny carbon nanotube bundles, some of which are only $\sim$10$^{-9}$ g in mass.

Specific heat and thermal conductivity are two of the most fundamental properties of the condensed matter. Over the past centuries, many methods have been developed for measuring these quantities. One noticeable category of the methods is the so called ``$3\omega$" method, which uses a narrow-band technique in detection, and therefore gives a relatively better signal-to-noise ratio, especially for the measurement on small specimens. In this method, either the specimen itself serves as a heater and at the mean time a temperature detector, if the specimen is electrically conductive and with a temperature-dependent electrical conductivity, or, for electrically non-conductive specimens, a metal strip is artificially deposited on the surface of the specimen to serve as the heater and detector. An {\it ac} current of the form $I_0{\rm sin}\omega t$ passing through the specimen or the metal strip creates a temperature fluctuation at an angular frequency of $2\omega$, and consequently, a resistance fluctuation at $2\omega$, which further generates a voltage fluctuation at $3\omega$. Corbino \cite{Corbino} is probably the first person noticed that the temperature fluctuation of an {\it ac} heated wire gives useful information about the thermal properties of the constituent material. Systematic investigations to the $3\omega$
method were carried out mainly during the 1960's \cite{Holland63,Gerlich65,Holland66}, and in the recent ten
years \cite{Cahill87,Birge87,Cahill89,Cahill90,Frank92,Jung92} which made this method a beautiful and practicable method. Nevertheless, we notice that in the previous works the heat-conduction equation were solved under the approximations either only for high-frequency limit \cite{Holland63,Gerlich65,Jung92}, or only for low-frequency limit \cite{Cahill87,Cahill89,Cahill90}. After these approximations one accordingly either loses the information of the thermal conductivity, or the specific heat of the specimen.

Here we present an explicit solution for the 1D heat-conduction equation. With this solution and together with the use of digital lock-in amplifier available nowadays, we are able to simultaneously obtain the specific heat and the thermal conductivity of a rod- or filament-like specimen through measuring the $3\omega$ voltage using a simple four-probe configuration. We have tested this method on Pt wire specimens. Correct values of specific heat, thermal conductivity, and Wiedemenn-Franz ratio, were obtained. With this method, we are able to measure the thermal properties of very tiny specimens such as carbon nanotubes, sometimes only 10$^{-9}$ g in mass.

In Section\ \ref{sec-solution} we will present an explicit solution of the heat-conduction equation. In Section\ \ref{sec-test} we will give the experimental test of the method on Pt wires (results on carbon nanotubes are already presented previously). A summary of the method will be in the last section.

\subsection{Solution of the 1D heat-conduction equation \\in the presence of an AC heating current}
\label{sec-solution}

We consider a uniform rod- or filament-like specimen in a four-probe measurement configuration as shown in Fig.\ \ref{configuration}. The two outside contacts are used to feed an {\it ac} current of the form $I_0{\rm sin}\omega t$ into the specimen, and the two inside ones are used to measure the voltage signals, just as in a standard electrical resistance measurement. The main differences from a pure electrical resistance measurement are (i) the part of the specimen in-between the two voltage contacts has to be suspended, to allow temperature variation along the specimen. (ii) All the contacts should be highly thermal-conductive, to heat-sink the specimen at these points to sapphire substrate. (iii) The specimen has to be maintained in high vacuum and the whole setup be heat-shielded to the substrate temperature, to preclude heat loss through air conduction and radiation. In such a configuration, the heat generation and diffusion along the specimen can be described by the following partial differential equation and the initial and boundary conditions:

\begin{figure}
	\centering
		\includegraphics[width=0.50\textwidth]{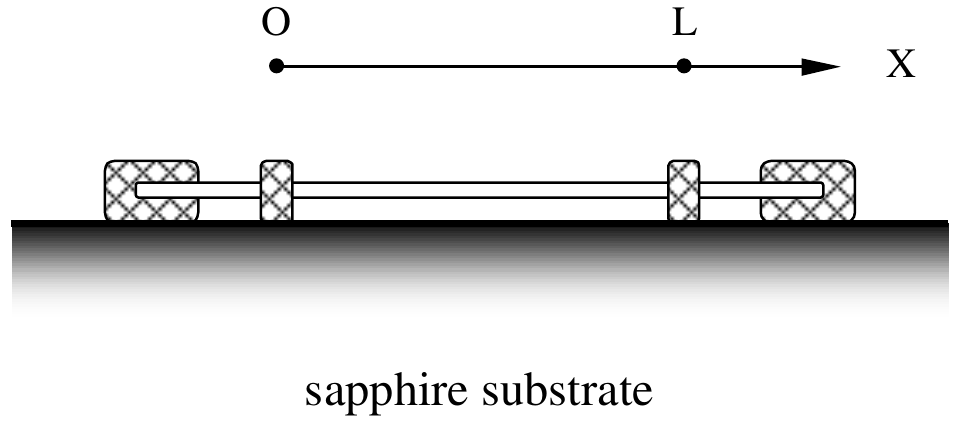}
\caption{Illustration of the four-probe configuration for measuring the specific heat and thermal conductivity of a rod- or filament-like specimen.} 
\label{configuration} 
\end{figure}

\begin{equation}
\rho C_p {\frac{\partial}{\partial t}}T(x, t)-\kappa {\frac{\partial^2}{\partial x^2}}T(x,t)
= \frac{I_0^2{\rm sin}^2\omega t}{LS} {\left[R+R'\left( T(x, t) - T_0\right) \right]}
\label{basic}
\end{equation}
\vspace{-0.5 cm}
\begin{equation}
\left\{ 
\begin{array}{l}
T(0, t)=T_0\\
T(L, t)=T_0\\
T(x, 0)=T_0 
\end{array}
\right.
\label{basic-c}
\end{equation}
where $C_p$, $\kappa$, $R$ and $\rho$ are the specific heat, thermal conductivity, electric resistance and mass density of the specimen, respectively, $R'=\frac{dR}{dT}$ at substrate temperature $T_0$, $L$, the length (between voltage contacts), and $S$, the cross section of the specimen. 

Setting $T(x, t)-T_0$ to $T(x, t)$ will not affect the solution but simplifies the equation and the initial and boundary conditions:
\begin{equation}
{\frac{\partial}{\partial t}}T(x, t) - a^2 {\frac{\partial^2}{\partial x^2}}T(x,t) 
-c~{\rm sin}^2\omega t~T(x, t) = b~{\rm sin}^2\omega t
\label{basicu}
\end{equation}
\vspace{-0.5 cm}
\begin{equation}
\left\{ 
\begin{array}{l}
T(0, t)=0\\
T(L, t)=0\\
T(x, 0)=0 
\end{array}
\right. 
\label{basicu-c}
\end{equation}
where $a^2=\frac{\kappa}{\rho C_p}$ is the thermal diffusivity, $b=\frac{I_0^2R}{\rho C_pLS}$, $c=\frac{I_0^2R'}{\rho C_pLS}$. 

Using the impulse theorem, $T(x, t)$ can be represented as the integral of the specimen's responses to the instant ``force", the right-side term of (\ref{basicu}), at each time interval:
\begin{equation}
T(x, t)=\int_0^t z(x, t; \tau)d\tau
\label{impulse}
\end{equation}
where $z(x, t; \tau)$ satisfies:
\begin{equation}
{\frac{\partial z}{\partial t}} - a^2 {\frac{\partial^2z}{\partial x^2}}
-c~{\rm sin}^2\omega t~z = 0
\label{z}
\end{equation}
\vspace{-0.5 cm}
\begin{equation}
\left\{ 
\begin{array}{l}
z(0, t)=0\\
z(L, t)=0\\
z(x, \tau+0)=b~{\rm sin}^2\omega\tau 
\end{array}
\right. 
\label{z-c}
\end{equation}

Expressing $z(x,t; \tau)$ with the Fourier series of the form:
\begin{equation}
z(x,t; \tau)=\sum_{n=1}^{\infty} U_n(t; \tau) {\rm sin}\frac{n\pi x}{L}
\label{z-int}
\end{equation}
and subtracting (\ref{z-int}) into (\ref{z}), we have
\begin{equation}
\sum_{n=1}^{\infty} \left[\frac{dU_n}{dt} + \left(\frac{n^2}{\gamma}-c~ {\rm sin}^2\omega t\right) U_n\right] {\rm sin}\frac{n\pi x}{L} = 0
\label{Un}
\end{equation}
where $\gamma\equiv (L/\pi a)^2$.

The term $c~{\rm sin}^2\omega t$ can be omitted if $n^2/\gamma\gg c$, or equivalently
\begin{equation}
\frac{I_0^2R'L}{n^2\pi^2\kappa S}\ll 1
\label{condition-1}
\end{equation}
This condition is usually well satisfied. For example, in a typical measurement $I_0$=10 mA, $R'$=0.1 $\Omega$/K, $L$=1 mm, $S$=10$^{-2}$ mm$^2$, $\kappa$=100 W/m~K, the left side of (\ref{condition-1}) is $\sim$ 10$^{-3}$. After omitting the $c~{\rm sin}^2\omega t$ term, the solution of the ordinary differential equation (\ref{Un}) is then:
\begin{equation}
U_n(t; \tau)=C_n(\tau) e^{-\frac{n^2}{\gamma}(t-\tau)}
\label{Un-s}
\end{equation}
where $C_n(\tau)$ can be determined using the initial condition in (\ref{z-c}), together with the relation $\sum_{n=1}^{\infty} \frac{2[1-(-1)^n]}{n\pi}{\rm sin}\frac{n\pi x}{L}=1$ for $0<x<L$:
\begin{equation}
C_n(\tau)=\frac{2b[1-(-1)^n]}{n\pi} {\rm sin}^2\omega\tau
\label{Cn}
\end{equation}

Using (\ref{Un-s}) and (\ref{Cn}), (\ref{z-int}) becomes:
\begin{equation}
z(x, t; \tau)=\sum_{n=1}^{\infty} {\rm sin}\frac{n\pi x}{L}~
\frac{2b[1-(-1)^n]}{n\pi}~ {\rm sin}^2\omega\tau ~
e^{-\frac{n^2}{\gamma}(t-\tau)}
\label{z-s}
\end{equation}

Subtract (\ref{z-int}), (\ref{z-s}) into (\ref{impulse}) and reset $T(x, t)$ to $T(x, t)-T_0$, we have:
\begin{equation}
T(x, t)-T_0=\sum_{n=1}^{\infty} {\rm sin}\frac{n\pi x}{L} ~
\frac{\gamma b [1-(-1)^n]}{n^3\pi}
\left[{1-\frac{{\rm cos}2\omega t+\frac{2\omega\gamma}{n^2}~{\rm sin}2\omega t}{1+\left(\frac{2\omega\gamma}{n}\right)^2}}\right]
\label{T-s}
\end{equation}
where we have already dropped off the exponential term which should fastly decay out with time. 

The total resistance of the specimen at temperature $T_0$ is:
\begin{equation}
R = \frac{1}{L}\int_{0}^{L}\left\{ R+R'[T(x, t)-T_0] \right\} dx
\label{Rtotal}
\end{equation}

Using (\ref{T-s}) and the relation 
$\int_{0}^{L} {\rm sin}\frac{n\pi x}{L} dx = [1-(-1)^n]\frac{L}{n\pi}$, resistance fluctuation with respect to $R$ at $T_0$ can be expressed as:
\begin{equation}
R_{fl}= \sum_{n=1}^{\infty} \frac{ \gamma bR' [1-(-1)^n]^2}{n^4\pi^2}
\left[ 1-\frac{{\rm cos}2\omega t+\frac{2\omega\gamma}{n^2}~{\rm sin}2\omega t}
{1+\left(\frac{2\omega\gamma}{n}\right)^2}\right]
\label{Rflucsum}
\end{equation}

Obviously, the $n=2$ term automatically vanishes. If only accounting the $n=1$ term, which introduces a relative error of the order $\sim 3^{-4}$, we have:
\begin{equation}
R_{fl}=\frac{4}{\pi^2}\gamma b R'
\left[ 1-\frac{{\rm sin}(2\omega t+\phi_0)}
{\sqrt{1+\left( 2\omega\gamma\right)^2}}\right]
\label{Rfluc}
\end{equation}
where $\phi_0$ satisfies:
\begin{equation}
{\rm tan} \phi_0=(2\omega\gamma)^{-1}
\label{phi0}
\end{equation}

The according voltage fluctuation in the presence of the {\it ac} current is $R_{fl}I_0{\rm sin}\omega t$. After dropping off the {\it dc} term and re-define a phase constant $\phi=\frac{\pi}{2}-\phi_0$, $ac$ voltage has the form:
\begin{equation} 
V_{3\omega}(t)=\frac {2I_0^3LRR'} 
{\pi^4\kappa S\sqrt{1+(2\omega \gamma)^2}}~{\rm sin}(3\omega t-\phi)  
\label{V3w} 
\end{equation} 
\begin{equation}
{\rm tan}\phi=2\omega \gamma 
\label{phi} 
\end{equation} 

If using rms values of voltage, $V_{3\omega}$, and rms values of current, $I$, as what lock-in gives, (\ref{V3w}) becomes:
\begin{equation} 
V_{3\omega}=\frac {4I^3LRR'} {\pi^4\kappa S\sqrt{1+(2\omega \gamma)^2}} 
\label{V3wrms} 
\end{equation} 

The following alternative form makes it more clearly how the 3$\omega$ signal depends on the dimensions of the specimen:
\begin{equation} 
V_{3\omega}=\frac {4I^3\rho_e\rho_e'} 
{\pi^4\kappa\sqrt{1+(2\omega \gamma)^2}}\left(\frac{L}{S}\right)^3
\label{V3wrmsdim} 
\end{equation} 
where $\rho_e$ is the electrical resistivity of the specimen, $\rho_e'=\frac{d\rho_e}{dT}$.

\begin{figure}
	\centering
		\includegraphics[width=0.50\textwidth]{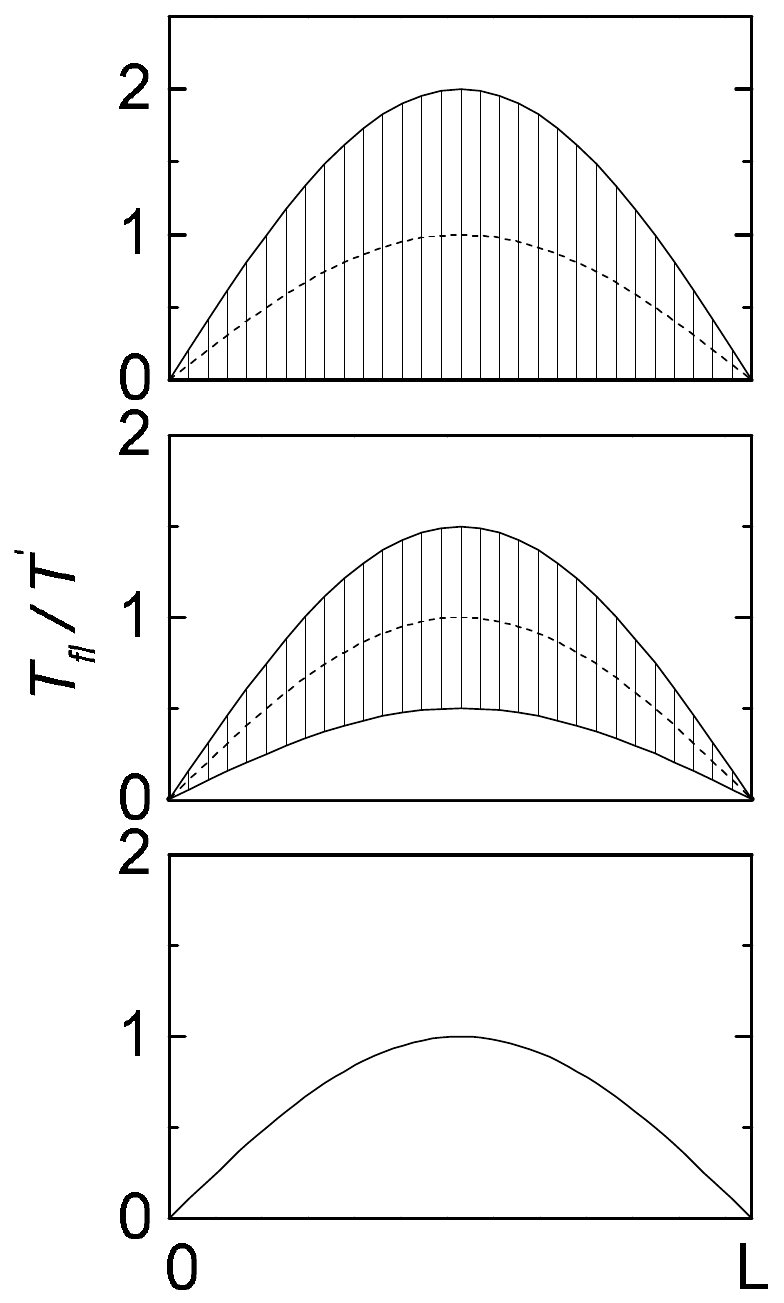}
\caption{Illustration of the temperature variation along the specimen under a driving {\it ac} current $I_0{\rm sin}\omega t$. The range of the variation is marked as shadowed area, which reaches the maximum at the limit $\omega\gamma\to 0$ and shrinks to a line as $\omega\gamma\to\infty$.} 
\label{tempfluc} 
\end{figure}

If accordingly only taking the $n=1$ term in (\ref{T-s}), the temperature fluctuation along the specimen is:
\begin{equation}
T_{fl}(x, t)=\frac{2\gamma b}{\pi}~{\rm sin}\frac{n\pi x}{L}
\left[ 1-\frac{{\rm sin}(2\omega t+\phi_0)}
{\sqrt{1+\left( 2\omega\gamma\right)^2}}\right]
\label{Tfluc}
\end{equation}
The result is illustrated in Fig.\ \ref{tempfluc}, where $T_{fl}^0=\pi/2\gamma b$. Obviously, temperature fluctuation reaches the maximum as $\omega\gamma\to 0$ (shown in Fig.\ \ref{tempfluc} (a)), whereas the temperature along the specimen saturates to a time-independent distribution as $\omega\gamma\to\infty$ (Fig.\ \ref{tempfluc} (c)).

At the high-frequency limit $\omega\gamma\gg 1$, {\it i.e.}, when the thermal wavelength $\lambda\ll L$, where $\lambda$ is defined as $\lambda=\sqrt{\frac{a^2}{2\omega}}$, (\ref{V3wrms}) becomes:
\begin{equation} 
V_{3\omega}=\frac {2I^3RR'} {\pi^2\omega\rho C_pLS}~~~~~(\omega\gamma\gg 1)
\label{V3wrmshf} 
\end{equation} 
which is the same as Holland's result \cite{Holland63} except for a slight difference in the numerical coefficients ({\it i.e.}, $\frac{2}{\pi^2}$ here {\it vs.} $\frac{1}{4}$ there). At this limit, however, one losses the information of thermal conductivity of the specimen. Oppositely, if taking the low-frequency limit, $\omega\gamma\ll 1$, or $\lambda\gg L$, one will only get the thermal conductivity of the specimen, but loss the information of specific heat, similar to what happened in Gahill's treatment for a two-dimensional heat diffusion problem \cite{Cahill90}. At the low-frequency limit, $V_{3\omega}$ is frequency-independent:
\begin{equation} 
V_{3\omega}=\frac {4I^3RR'L} {\pi^4\kappa S}~~~~~~(\omega\gamma\ll 1)
\label{V3wrmslf} 
\end{equation} 

In above derivation we have neglected the radial heat loss through radiation. Such heat loss per unit length from a cylindrical rod of diameter $D$ can be expressed as:
\begin{equation}
W(x, t) =\pi\sigma D\left[ T^4(x, t)-T_0^4\right]
\approx 4\pi\sigma DT_0^3\left[ T(x, t)-T_0\right]
\label{heatloss} 
\end{equation} 

Equation (\ref{basicu}) and (\ref{basicu-c}) can be re-written as:
\begin{equation}
{\frac{\partial}{\partial t}}T(x, t) - a^2 {\frac{\partial^2}
{\partial x^2}}T(x,t) 
\left( g-c~{\rm sin}^2\omega t\right) T(x, t) = b~{\rm sin}^2\omega t
\label{basicur}
\end{equation}
\vspace{-0.5 cm}
\begin{equation}
\left\{ 
\begin{array}{l}
T(0, t)=0\\
T(L, t)=0\\
T(x, 0)=0 
\end{array}
\right. 
\label{basicur-c}
\end{equation}
where $g=\frac{16\sigma T_0^3}{\rho C_pD}$. (\ref{Un}) now becomes:
\begin{equation}
\sum_{n=1}^{\infty} \left[\frac{dU_n}{dt} + \left(\frac{n^2}{\gamma}+g-c~ {\rm sin}^2\omega t\right) U_n\right] {\rm sin}\frac{n\pi x}{L} = 0
\label{Unr}
\end{equation}

After replacing $\frac{n^2}{\gamma}+g$ with $\frac{n^2}{\gamma '}$, the final solution of (\ref{Unr}) is the same as (\ref{V3w}) and (\ref{phi}), except that now $\gamma$ in (\ref{V3w}) and (\ref{phi}) should be replaced with $\gamma '$. In the approximation of only taking the $n=1$ term, we have:
$\gamma ' = \frac{\gamma}{1+g\gamma}$.
Obviously, radiation heat loss can be neglected if 
\begin{equation}
g\gamma\ll 1
\label{condition-2}
\end{equation}
For cylindrical rod, (\ref{condition-2}) becomes: $\frac{16\sigma T_0^3L^2}{\pi^2\kappa D}\ll 1$. Shortening the length or increasing the diameter of the rod will help to eliminate the effect of heat loss through radiation.

Similar discussion can also be applied to account for the heat loss by remnant air. But for detailed calculation one needs to know the thermal resistance of the specimen at the surface.

\subsection{Experimental test}
\label{sec-test}

The test of the above solution was carried out on two kinds of specimens: Pt wires and bundles of multiwall carbon nanotubes. The electrical resistance of the former has a positive temperature coefficient, whereas the latter a negative coefficient. Within the reasonable parameter ranges of frequency and current (we will explain it later), we do find that the $V_{3\omega}$ is proportional to $I^3/\sqrt{1+(2\omega\gamma)^2}$, and the phase angle ${\rm tan}\phi$ is proportional to the angular frequency. For Pt specimen, the yielded specific heat and thermal conductivity, as well as the Wiedemenn-Franz ratio, do agree with the standard data over the entire temperature range measured.

\begin{figure}
	\centering
		\includegraphics[width=0.70\textwidth]{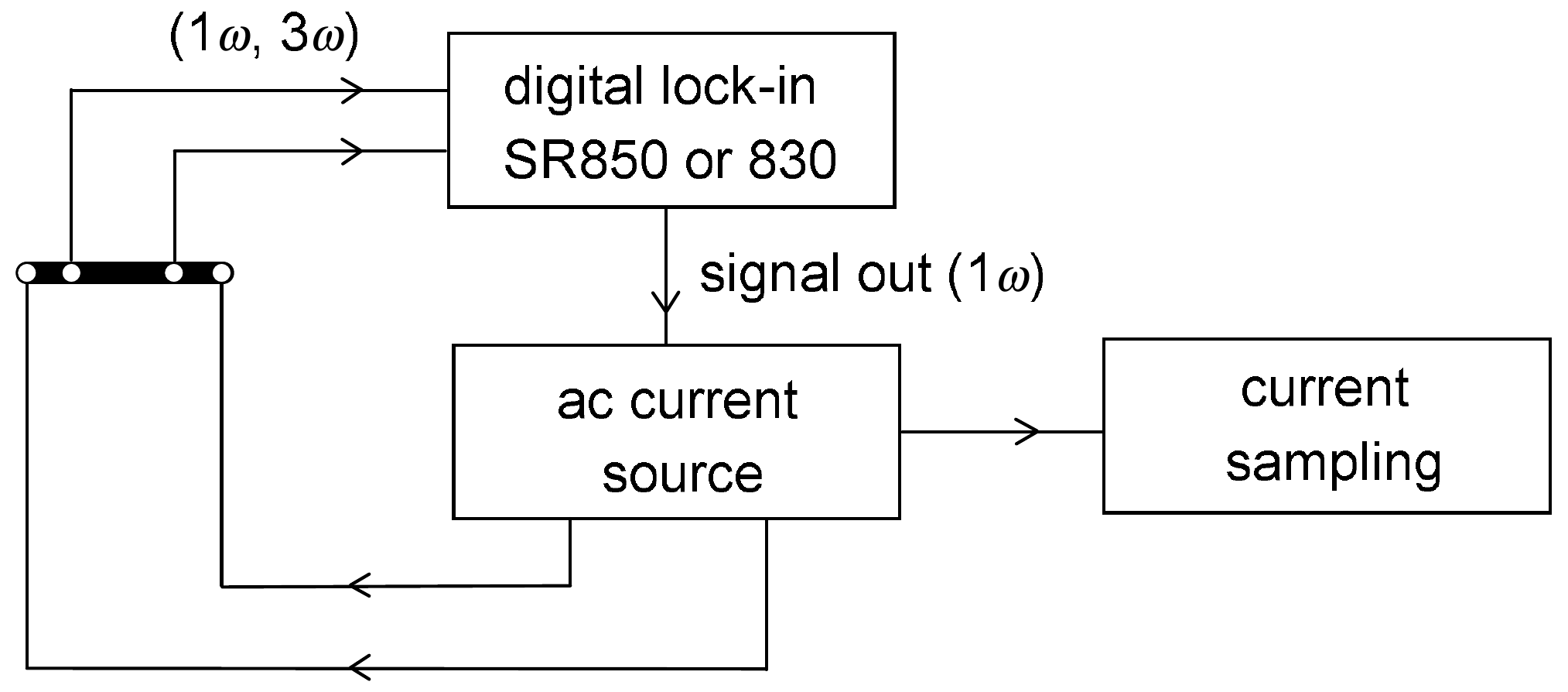}
\caption{Block diagram of the measurement. We choose the digital lock-in amplifier SR850 or 830 to measure the 3$\omega$ voltage. The reference-out of the lock-in is boosted to be an {\it ac} current by a simple circuit (lower panel), and is then applied to the specimen. The feed-back resistor R$^*$ should be those with very low temperature coefficient.} 
\label{block} 
\end{figure}

We use a digital lock-in amplifier SR850 to measure the 3$\omega$ voltage signal. The block diagram of measurement is shown in Fig.\ \ref{block}. We turned off all the filters of the lock-in during the measurement, and used the {\it dc} coupling input mode. Before measuring the 3$\omega$ signal, the phase of the lock-in amplifier was adjusted to zero according to the 1$\omega$ signal. We use a simple electronic circuit to booster the 1$\omega$ reference-out voltage into an {\it ac} current feeding to the specimen. The 3$\omega$ component of the current is 10$^{-4}$ or less comparing with the 1$\omega$ component, as checked by an HP89410A spectrum analyzer. Because the 3$\omega$ voltage signal is deeply buried in the 1$\omega$ voltage signal of many order of magnitude higher in amplitude, high dynamic reservation of the lock-in is usually required. We keep the dynamic reservation unchanged relative to the total magnification of the lock-in amplifier during the entire measurement. 

There are two ways to perform the measurement. In the first way the substrate of the specimen is maintained to a few fixed temperatures. At each temperature the frequency dependence of the $V_{3\omega}$ is measured. In this way one can check whether the $I^3$ and the $1/\sqrt{1+(2\omega\gamma)^2}$ dependencies of $V_{3\omega}$ , as well as the ${\rm tan}\phi=2\omega\gamma$ relation, are held. 

Because $V_{3\omega}\propto I^3$, one will get much larger signal if using larger $I$. However, there are two reasons that prevent us from using inadequately large $I$. Firstly, excessive heat accumulation on the specimen will create significantly large temperature gradient at the silver paste contacts, which will violate the boundary condition in (\ref{basic-c}). Secondly, radiation heat loss will increase significantly as the temperature modulation is increased. In all the cases the expected relation $V_{3\omega}\propto I^3$ will no longer be held. On the other hand, the relation will also be violated if using too small $I$, since in the latter case $V_{3\omega}$ could be comparable to, or even smaller than, the spurious 3$\omega$ signal from the current and other sources. In our measurement the total heating power at each measuring temperatures was maintained to be approximately equal to the value of the specimen's thermal conductivity, {\it i.e.}, keeping the temperature modulation to be around 1 K along the specimen. 

In the second way of measurement, the temperature of the substrate is slowly ramped at a fixed rate, meanwhile the working frequency of the lock-in amplifier is switched between a few set values. The whole process, including the temperature ramping rate, the amplitude of the heating current, the frequency set-points, are all controlled by a personal computer. 

For Pt specimen, we chose a wire of diameter $D=20$ $\mu$m and length $L=8$ mm. We found that 
$\gamma$ of the specimen varied from 0.005 s$^{-1}$ at 10 K to $\sim$ 0.2 s$^{-1}$ at room temperature, so that the working frequencies were chosen to be between 1 to 80 Hz, in order to cover the most featured frequency range of the functional form $1/\sqrt{1+(2\omega\gamma)^2}$. 

\begin{figure}
	\centering
		\includegraphics[width=0.70\textwidth]{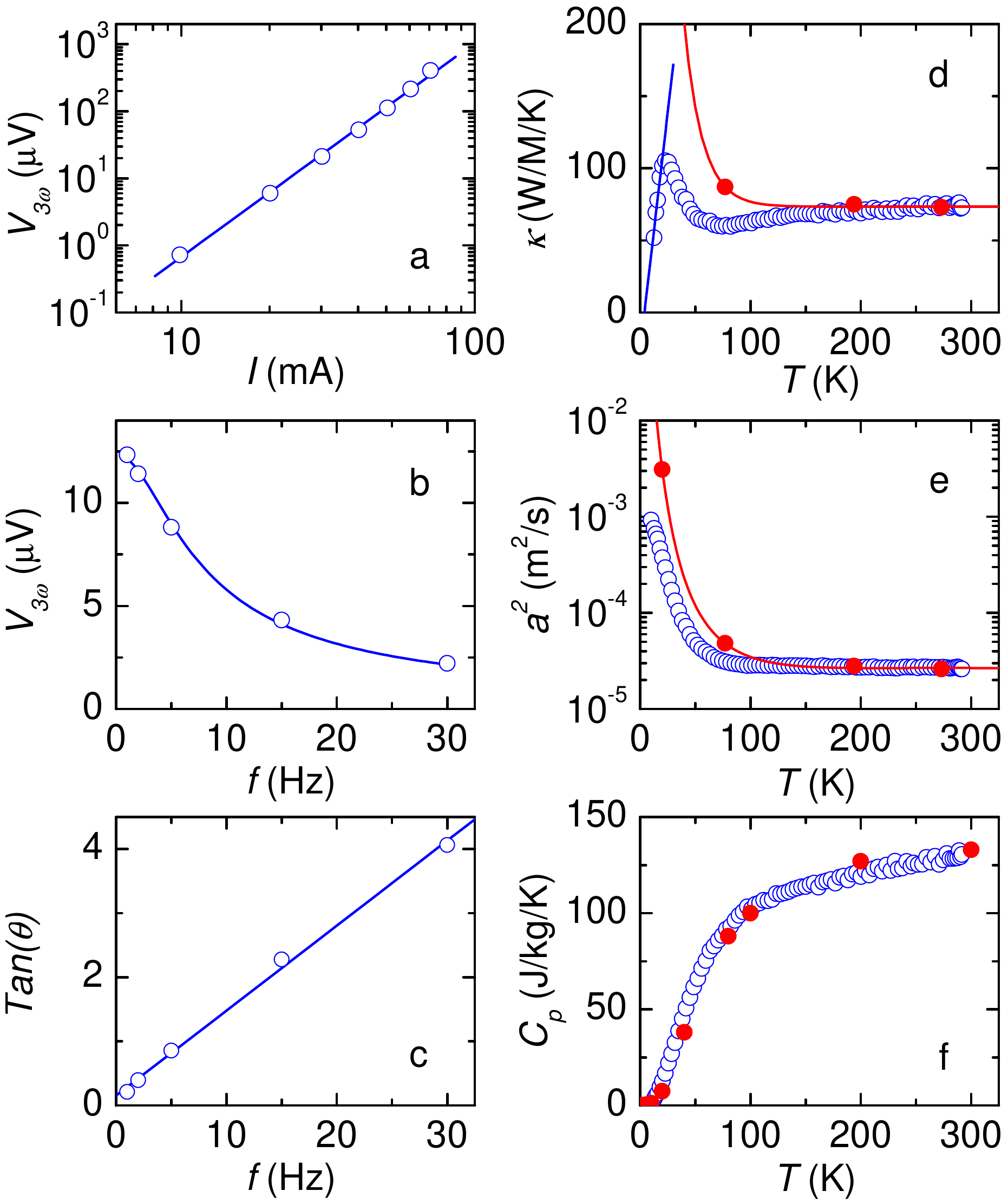}
		\caption{Experimental test of the 3$\omega$ method on Pt wire of 20 ${\rm\mu m}$ in diameter. (a) Current dependence of the $V_{3\omega}$. The open circles are the measured data, and the solid line is the predicted relation $V_{3\omega}\propto I^3$. (b) Frequency dependence of the $V_{3\omega}$ (open circles). The solid line is the predicted relation $V_{3\omega} \propto 1/\sqrt{1+(2\omega\gamma)^2}$. (c) Frequency dependence of the phase angle of the $V_{3\omega}$ (open circles). The solid line is again the predicted relation ${\rm tan}\phi\propto\omega\gamma$. The thermal conductivity $\kappa$, thermal diffusivity $a^2$, and specifc heat $C_p$ of the Pt specimen are plotted as open circles in figures (d), (e), and (f), respectively. Also shown as solid circles in these figures are the standard data of Pt \protect{\cite{C-of-Pt}}. The difference in $\kappa$ and $a^2$ between our measured data and the standard data should reflect the difference in purity and/or structural perfection between the Pt specimens of different sources.}
\label{Pt} 
\end{figure}

Shown in Fig.\ \ref{Pt} (a) is the current dependence of the $V_{3\omega}$ at 22K, demonstrating an $I^3$-law in a mediate current range. Figure\ \ref{Pt} (b) and (c) show the frequency dependencies of the amplitude and the phase angle of the $V_{3\omega}$ signal at 22 K, respectively. It can be seen that the data (open circles) agree well with the predicted functional forms (solid lines). By fitting the data in Fig.\ \ref{Pt} (b)
to the functional form of (\ref{V3wrms}), we get the information of thermal conductivity $\kappa$ (Fig.\ \ref{Pt} (d), open circles) and the time constant $\gamma$. Note that the $\gamma$ thus obtained is in fairly good agreement with what the phase angle reveals {\it via} Eq. (\ref{phi}). Thermal diffusivity $a^2$ and specific heat $C_p$ of the specimen can be obtained {\it via} relation $\gamma\equiv (L/\pi a)^2$ and $a^2=\frac{\kappa}{\rho C_p}$. The results are shown in Fig.\ \ref{Pt} (e) and (f) as open circles. $C_p$ thus obtained agrees well with the standard data for Pt \cite{C-of-Pt} (the solid circles in Fig.\ \ref{Pt} (f)).

\begin{figure}
	\centering
		\includegraphics[width=0.50\textwidth]{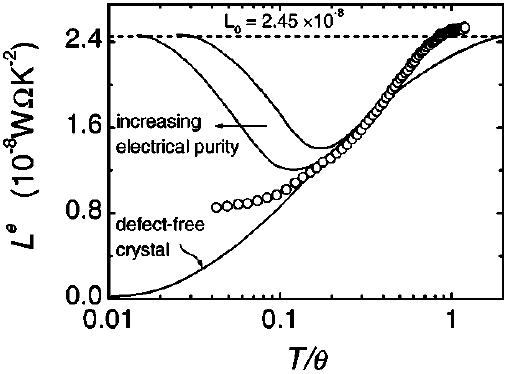}
		\caption{The Wiedemenn-Franz ratio $L^e$ of our Pt specimen compared with that of usual metals with different purity. The result indicates that our Pt wire specimen is pure but not totally defect-free. The room-temperature Wiedemenn-Franz ratio of our Pt specimen is about $2.52\times 10^{-8} {\rm W\Omega/K^2}$, in perfect agreement with what reported in literature \protect{\cite{L-of-Pt}}, but is slightly larger than that of the free-electron Lorenz number $L_0=2.45\times 10^{-8} {\rm W\Omega/K^2}$ (the dashed line). Note that for Pt its Debye temperature $\theta$ is 240 K.}
\label{Lorenz} 
\end{figure}

We notice that the measured thermal conductivity of the Pt wire only show a small peak at low temperatures, far insignificant than what is expected for high purity Pt material. Such difference is possible since $\kappa$ depends largely on the purity, structural perfection, and even the dimensions of the specimen. We believed that the $\kappa$ data we obtained reflect the true thermal conductivity of our Pt specimen, since the Wiedemenn-Franz ratio deduced from the $\kappa$ and the electrical resistivity, or more directly, from the thermal conductance and the electrical resistance, fits well to the case of pure but not totally defect-free metal samples in the whole temperature range measured \cite{L-of-Pt}, as shown in Fig.\ \ref{Lorenz}. The absolute value of the Wiedemenn-Franz ratio is found to be $\sim 2.53 \times 10^{-8} {\rm W\Omega/K^2}$ at 290 K, slightly larger than that of the free-electron Lorenz number $2.45\times 10^{-8} {\rm W\Omega/K^2}$, which is also in perfect agreement with the previously reported value in literature \cite{L-of-Pt}. 

\begin{figure}
	\centering
		\includegraphics[width=0.50\textwidth]{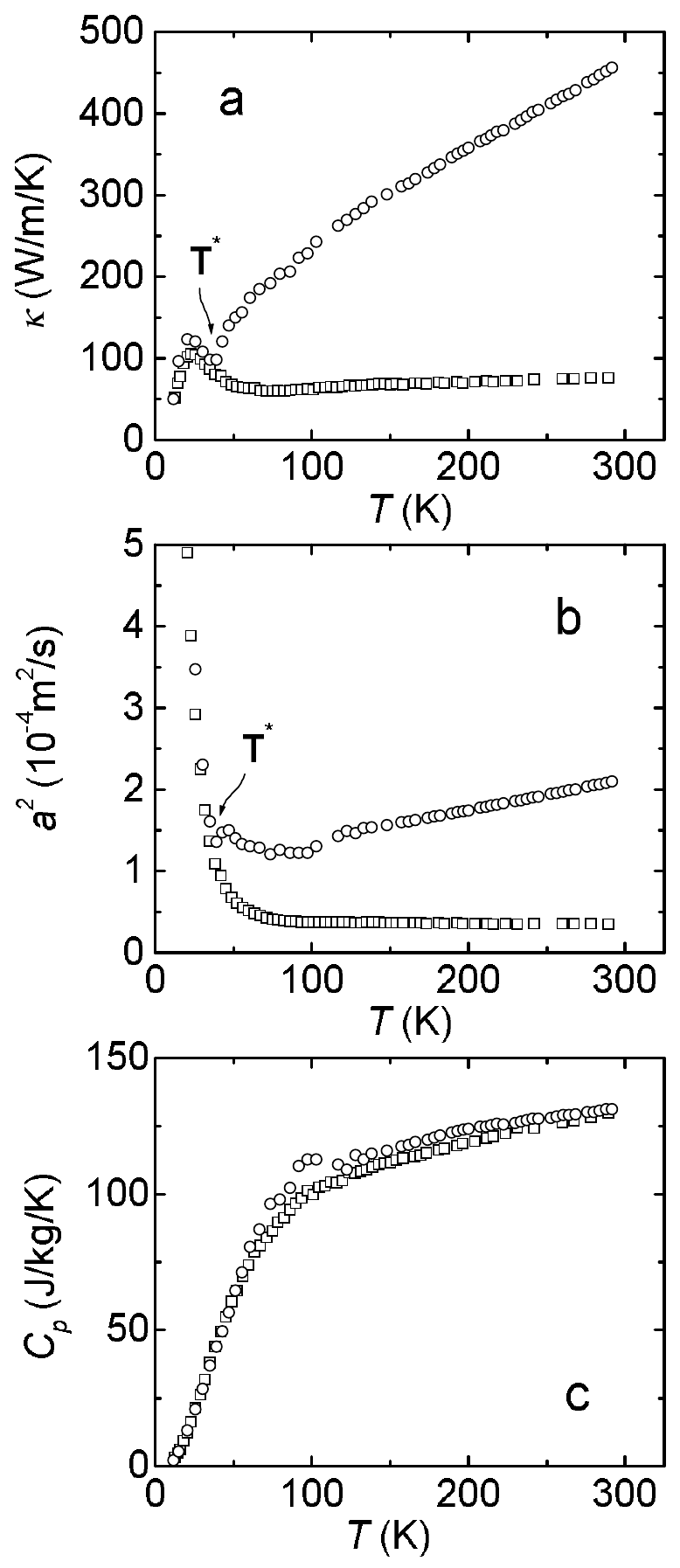}
		\caption{Effect of radial heat loss through air conduction. Open squares represent the case where such heat loss is absent. In one run of the measurement, radial heat loss was triggered on above $T^*$ (marked in the figure) by distroying the vaccum of the system. The heat loss results in spurious larger thermal conductivity and diffusivity of the specimen (open circles in (a) and (b)). Nevertheless, the specific heat deduced from the $\kappa$ and $a^2$ data is relatively insensitive to such heat loss (Fig. (c)).}
\label{PtP} 
\end{figure}

Let us now examine the effect of radial heat loss through air conduction. Different sets of data in Figs.\ \ref{PtP} (a), (b) and (c) were taken at different vacuum pressures. Open squares correspond to the data taking in high vacuum where slight changes ({\it e.g.}, a factor of 2) in vacuum pressure does not affect the results, {\it i.e.}, heat loss through air is absent. In poor vacuum, however, such heat loss results in spurious larger thermal conductivity and diffusivity of the specimen (open circles in Figs.\ \ref{PtP} (a) and (b)). Nevertheless, it is interesting that the specific heat deduced from the $\kappa$ and $a^2$ data is relatively insensitive to such heat loss (Figs.\ \ref{PtP} (c)). 

After all, let us check if the approximation conditions (\ref{condition-1}) and (\ref{condition-2}) are satisfied. We find that
$\frac{I_0^2R'L}{n^2\pi^2\kappa S}\sim 10^{-5}$ (taking $n=1$). Therefore, (\ref{condition-1}) is well satisfied. For radiation heat loss, we find that $g\approx$ 0.22 s$^{-1}$ even at 300 K, and $\gamma$ (actually, 
$\gamma '$) deduced from the measurement is about 0.2 s. Therefore, $g\gamma\approx$ 0.044, which means that omitting the radiation heat loss will cause about 5\% of error for a 20 $\mu$m diameter Pt wire specimen with length of 8 mm.

We have also applied the 3$\omega$ method to measure the $\kappa$ and $C_p$ of multiwall carbon nanotube bundles who have a negative temperature coefficient in electrical resistivity. The bundles are about $\sim$1 mm long and with a diameter ranging from $\sim$1 $\mu$m up to $\sim$100 $\mu$m. For these specimens, no $C_p$ and $\kappa$ data of other sources are available for comparison. Nevertheless, the frequency dependencies of $V_{3\omega}$ and ${\rm tan}\phi$ obtained on the specimens are all in good agreement with the predicted functional forms (see Fig.\ \ref{setup}), which guarantees the reliability of the deduced $\kappa$ and $C_p$. For the specimens in this measurement,  
$\frac{I_0^2R'L}{\pi^2\kappa S}$ is below 10$^{-3}$ above $\sim$ 60 K and is about 0.08 at 10 K, and $g\gamma$ is below 5$\times 10^{-3}$ in the whole temperature range, both of which satisfy the conditions (\ref{condition-1}) and (\ref{condition-2}), respectively. For a 1 $\mu$m diametered specimen, the mass is about 10$^{-9}$ g, which is far less than the minimum amount of mass required in any other kinds of $C_p$ measurement. 

\subsection{Summary}

We have presented an improved 3$\omega$ method after explicitly solved the heat-conduction equation under some reasonable approximations. This method can simultaneously measure the thermal conductivity and the specific heat of electrically conducting materials in a way similar to, and as simple as usual electrical resistivity measurement.

\putbib[ch3-ref]
\end{bibunit}


\chapter{Tunneling Spectroscopy}

\begin{bibunit}[unsrt]

\setcounter{section}{0}

\section{Motivation}

Coulomb blockade (CB) has been intensively studied in a
multijunction configuration, in which electron tunnel rates from
the environment to a capacitively isolated "island" are blocked by
the e-e interaction if the thermal fluctuation is below the
charging energy $E_c = e^{2}/2C$ and the quantum fluctuation is
suppressed with sufficiently large tunnel resistance $R_t \gg R_Q
= h/2e^2$. In the case of a single-junction circuit, the
understanding of CB is less straightforward. The Coulomb gap,
supposed washed out for the case of low-impedance environments,
should only be established if the environmental impendence exceeds
$R_Q$.

In singlewalled carbon nanotubes (SWNTs), the reduced geometry
gives rise to strong e-e interaction. Indeed, CB oscillations and
evidence of Luttinger liquid (LL) have been
observed \cite{Bockrath}. In contrast to SWNTs, in which only two
conductance channels are available for current transport, MWNTs
with diameter in the range of 20--40 nm have several tens of
conductance channels. The energy separation of the quantized
subbands, given by $\Delta E = \hbar \upsilon_f/d$, is about
13--26 meV taken Fermi velocity as $\upsilon_f = 8 \times 10^{5}$
m/s. This value is about an order of magnitude smaller than that
of SWNTs. Experiments indicate that MWNTs are considerably hole
doped, thus a large number of subbands in the order of ten are
occupied. The observation of weak localization \cite{Langer},
electron phase interference effect \cite{Bachtold2} and universal
conductance fluctuations \cite{Langer} support the view that low
frequency conductance in MWNTs is contributed mostly by the
outmost graphene shell and is characterized by 2D diffusive
transport. Besides the phase interference correlated effects,
strong e-e interaction has also been observed in MWNTs. There have
been several tunneling spectroscopy observations of a pronounced
zero-bias suppression of tunneling density of states (TDOS) \cite{Haruyama, 
Tarkiainen,Bachtold}. Moreover, the TDOS shows a power law asymptotics, i.e.,
$\nu(E) \sim E^{\alpha}$, which resembles the case of a LL. It is
noteworthy that in the Environmental Quantum Fluctuation (EQF) theory, 
for a single tunnel junction
coupled to high-impedance transmission lines, such a scaling
behavior is also predicted in the limit of many parallel
transmission modes \cite{Ingold}. The physical origin of
these power laws is the linear dispersion of bosonic excitations
that are characteristic both for LL, which is strictly 1D
ballistic conductor, and a single tunnel junction connected to a
3D disordered conductor. In the latter case the quasi particle
tunneling is suppressed at $V \ll e/2C$, therefore the charge is
transported with 1D plasmon modes. The facts that MWNTs have many
conductance modes, as well as the observation of a crossover from
power law to Ohmic behavior at higher voltages \cite{Tarkiainen},
suggests that EQF theory is more appropriate to describe the
observed TDOS renormalization. Most of these measurements were
done in multi-junction configurations. Therefore a single tunnel
junction measurement is needed to further clarify this issue.

\section{Experimental}

In this chapter, millimeter-long CVD-grown MWNTs \cite{Pan} are
measured by a cross-junction method. The MWNTs samples are
composed of loosely entangled nanotubes that are roughly parallel
to each other and extend up to 2 mm long. The single tunnel
junctions are formed by crossing a very thin ($<1$ $\mu$m) MWNTs
bundle on a narrow strip of metal wire fabricated on an insulating
substrate. We explored different electrode materials including Au,
Cu, Sn, and Al. In the case of Sn and Al, a small magnetic field
was applied to suppress the Superconducting state below Tc. No
obvious change of device characteristics caused by the choice of
metals was observed. With such a cross-junction configuration the
measured conductance is only contributed by the tunnel junction
itself, since the current is passed along one arm of the
MWNT/metal and voltage is measured along the other non-current
carrying arm. Thus the device can be understood as a small number
of single junctions in parallel. Despite the simplicity of the
fabrication method, we found that the devices made are very stable
and sustain several cooling cycles without apparent change of
characteristics. More than 20 samples we measured all yield strong
zero bias suppression of TDOS.

\begin{figure}
	\centering
		\includegraphics[width=0.90\textwidth]{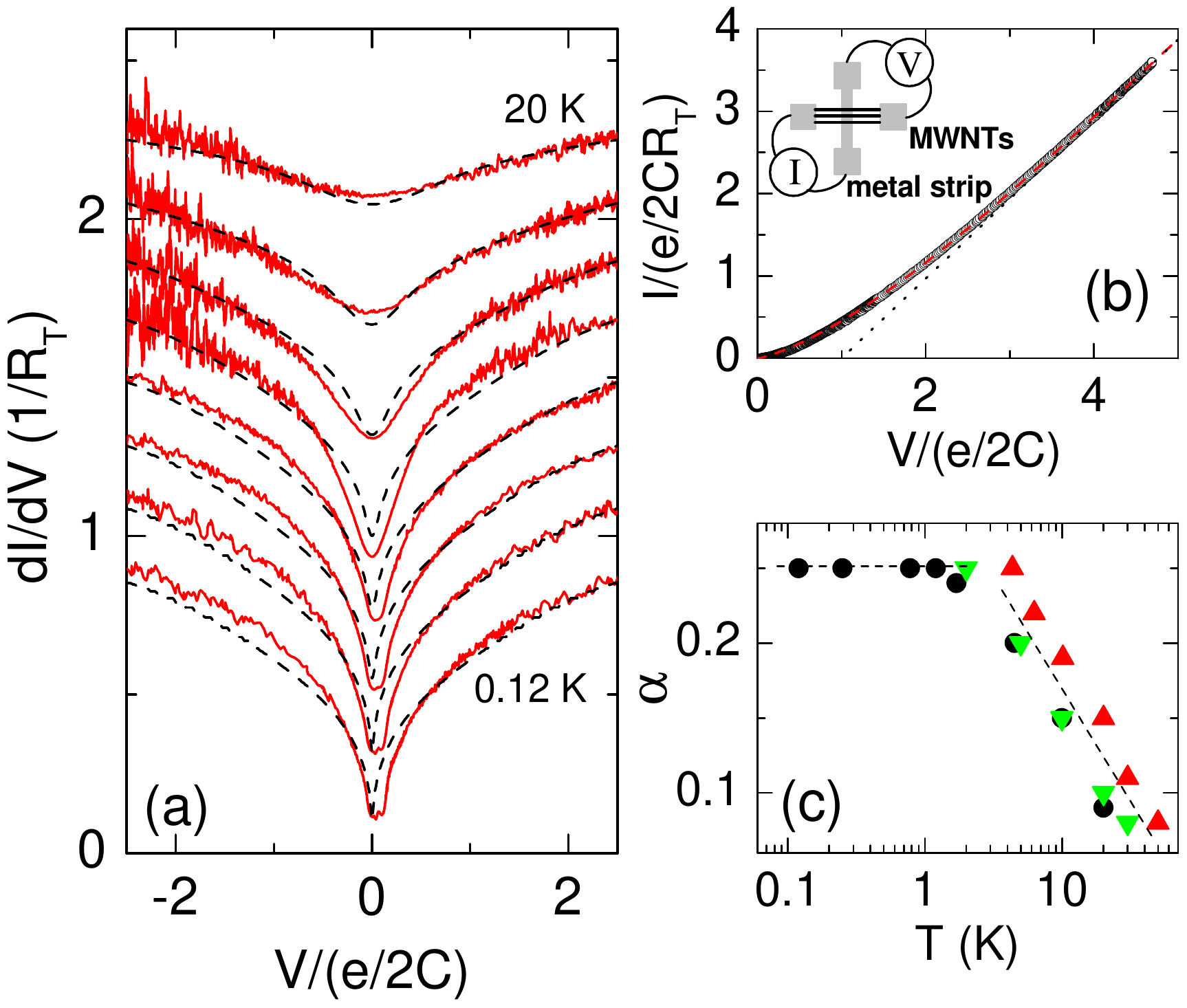}
	\caption{(a) $dI/dV$ as a function of $V$ in
dimensionless units measured at $T$ = 0.12, 0.25, 0.78, 1.2, 1.7,
4.5, 10, 20 K (sample 990530s6). Dashed lines are the fits of EQF
theory. (b) DC current simultaneously measured at 0.12 K. Dashed
line is the numerical fit. Dotted line shows a Coulomb offset of
$e/2C$. (c) Temperature dependence of the exponent $\alpha$ for
three samples ($\bigcirc$: 990530s6; $\bigtriangleup$: 990316;
$\bigtriangledown$: 990320).}
\label{dIdV}
\end{figure}

The current I(V) as well as $G \equiv dI/dV$, are calculated by a
golden-rule approach incorporating the environmental influence by
$P(E)$, the probability for a tunneling electron to lose energy to
the environment. $P(E)$ can be calculated from its Fourier
transform, the phase correlation function $J(t)$. In the
transmission-line model, $J(t)$ is determined by the total
environmental impedence
\begin{equation}
Z_t (\omega) = (i\omega C + Z^{-1}(\omega))^{-1}
\end{equation}
where $Z(\omega)$ is the external
environmental impedance. We applied the method in Ref.\ \cite{
Ingold2} to evaluate $P(E)$ from an integral equation without the
need of going to the time domain. The I--V characteristics are then
calculated for finite temperatures and arbitrary junction
impedance. The parameters that appear in the numerical calculation
are the damping strength 
\begin{equation}
\alpha = Z(0)/R_Q, 
\end{equation}
the inverse relaxation time $\omega_{RC}$, and the quality factor 
\begin{equation}
Q = \omega_{RC} / \omega_S 
\end{equation}
with the resonance frequency of the undamped circuit 
$\omega_S = (LC)^{-1/2}$. It is noted that the
quality factor $Q$ only plays a minor role as an additional
adjustable parameter. Therefore it is not important to the fit,
and is set always to unit.

\section{Discussion}

We then tried to fit our experimental data with the EQF theory.
Since $E_c$ and $R_t$ are determined by the high-voltage data, the
isolation resistance $R_{iso} = Z(0)$ is the only adjustable
parameter in our fitting. In contrast to the case of Ref.\ \cite{Zheng}, where $R_{iso}$ is formed by ideal Ohmic and
temperature-independent resistors, $R_{iso}$ in our case are
provided by the resistive impedance of the MWNTs themselves, which
should be temperature-dependent. From the fitting, we do see an
evolution of $\alpha$ from 0.1 at 20 K to 0.25 at 1--4 K and then
$\alpha$ saturates (Fig.\ \ref{dIdV}c). Note that dimensionless units are
used with $G$ normalized by 1/$R_t$ and voltage normalized by
$e/2C$. Therefore the number of the single tunnel junctions has no
effect on the fitting and results on different samples can be
directly compared. We found that for different samples with
scattered characteristics (see Table\ \ref{table1}), at low temperatures the
exponent $\alpha$ reaches an universal value of 0.25--0.35, which
agrees with previous multi-junction measurements \cite{Tarkiainen,
Bachtold}. It yields $R_{iso}$ = 3.3--4.6 k$\Omega$ that is
roughly a constant for different samples. This coincidence is not
accidental but reflects the intrinsic electrodynamic modes of the
MWNTs, which can be modeled as an ideal resistive LC transmission
line: 
\begin{equation}
R = (L'/C')^{1/2}
\end{equation}
with the kinetic inductance $L' = R_Q/2N\upsilon_f$  estimated as 
$\approx 1$ nH/$\mu$m for $N\approx$ 10--20 modes and the capacitance 
$C' \approx$ 20--30 aF/$\mu$m. Therefore, the ``environment'' with respect to the
single-junctions in our devices is provided by MWNTs themselves,
not the external circuits.

\begin{table}
\begin{center}
\caption{\label{tab:table1}A partial list of characteristics of the samples.}
\label{table1}
\begin{threeparttable}
\begin{tabular}{cccccccc}
Sample &$R_t$ (k$\Omega$)&$E_c$ (eV) &$C$ (aF) &$R_{iso}$ (k$\Omega$)&$\alpha$\\
\hline
990316& 18.4 & 0.019 & 4.2 &3.3\tnote{a} & 0.25\tnote{a} \\
990320& 48.3 & 0.014 & 5.7 &3.3\tnote{a} & 0.25\tnote{a} \\
990530s6& 8.6 & 0.01 & 8.0 &3.3\tnote{a} & 0.25\tnote{a} \\
990530s1& 10.0 & 0.02 & 4.0 &4.6\tnote{a} & 0.35\tnote{a} \\
990202& 9.1 & 0.004 & 20.0 &3.5\tnote{a} & 0.27\tnote{a} \\
\end{tabular}
\begin{tablenotes}
\item[a] Value acquired at $T$ = 1.2--4.3 K.
\end{tablenotes}
\end{threeparttable}
\end{center}
\end{table}

\begin{figure}
	\centering
		\includegraphics[width=0.70\textwidth]{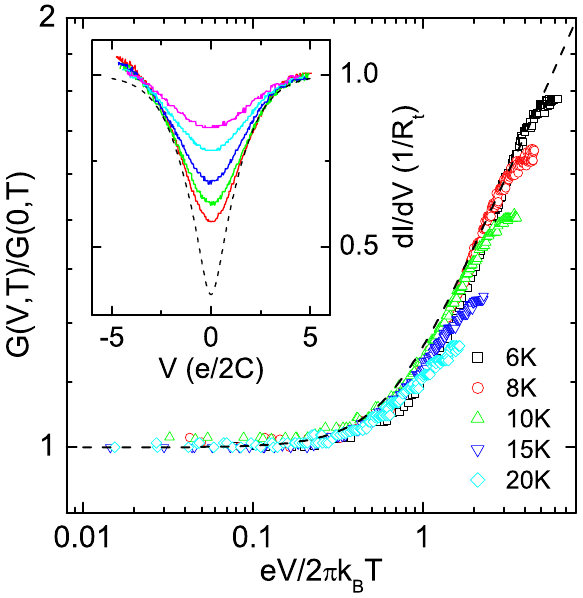}
	\caption{Scaled conductance $G(V,T)/G(0,T)$ of
another sample (990202). The inset shows the original $dI/dV$ vs
$V$ with the dashed line calculated with exponent $\alpha =
0.27$.}
\label{scaled}
\end{figure}

As mentioned previously, for an isolated single tunnel junction
with many parallel transport modes, the EQF theory predicts a
power-law asymptotics: 
\begin{equation}
G(V,T)/G(0,T) \equiv f(V/T) = \lvert\Gamma[\frac{\alpha}{2} + \frac{ieV}{2\pi
k_BT}]/\Gamma[\frac{\alpha}{2}]\Gamma[1 + \frac{ieV}{2\pi k_BT}]\rvert^{2}
\end{equation}
where $G(0,T)$ is the zero-bias conductance \cite{Zheng}. For $eV/2\pi k_B T \gg 1$, a voltage
power law $G(V,T) \sim V^\alpha$ is expected. Note that in the
above scaling function, exponent $\alpha$ is the only adjustable
parameter. In a two-junction configuration, an additional parameter was introduced
to take into account the voltage division between the two tunnel junctions \cite{Bockrath}. Seen in Fig.\ \ref{scaled}, we do observe such a scaling behavior.
Here $f(V/T)$ represented by solid line is calculated taken
$\alpha = 0.27$. The inset shows the original data in which the
dashed line is calculated numerically using the same exponent
$\alpha = 0.27$. 

CB and Coulomb gap effects in MWNTs are recently
described with microscopic theories in Refs. [14,15]. In
the 2D diffusive regime, the exponent for tunneling into
the bulk of a MWNT is 
\begin{equation}
\alpha=(R/h\nu_{0}D)ln(1+\nu_{0}U_{0})
\end{equation}
Here \textit{R} is the tube radius, $U_{0}$ is 
the intratube Coulomb interaction, $D=\upsilon^{2}_{f}\tau/2$
is the charge diffusivity, and the ``bare'' DOS $\nu_{0}=N/4h\upsilon_{f}$ 
with $N \approx 20$. If we use $\upsilon_{f}\tau \approx 60$ nm 
from the magnetoresistance data \cite{Yi-unpublished} and taking
$R \approx 20$ nm, then under the condition of $U_{0}/h\upsilon_{f}\sim 1$ the
above formula yields $\alpha \approx 0.25$, which agrees with our
experiment.

\section{Fano resonance}

Besides the voltage power-law phenomenon, which is
characteristic of strong Coulomb interactions, another
energy scale ``the discrete energy levels due to electron
confinement'' emerges as temperature decreases, and it
modifies the tunneling spectra. Unlike the case of a two-junction
configuration, where the electrons are confined
by the two contacts if the nanotube is clean enough, in a
single-junction configuration the electrons can be confined by disorders 
when the MWNTs are ``dirty'', such
that the impedance of the local environment of the junction
is larger than $R_Q$. A stacking mismatch between
adjacent walls and other structural imperfections are
possible sources of disorders in MWNTs, resulting in
discrete energy levels.

\begin{figure}
	\centering
		\includegraphics[width=0.70\textwidth]{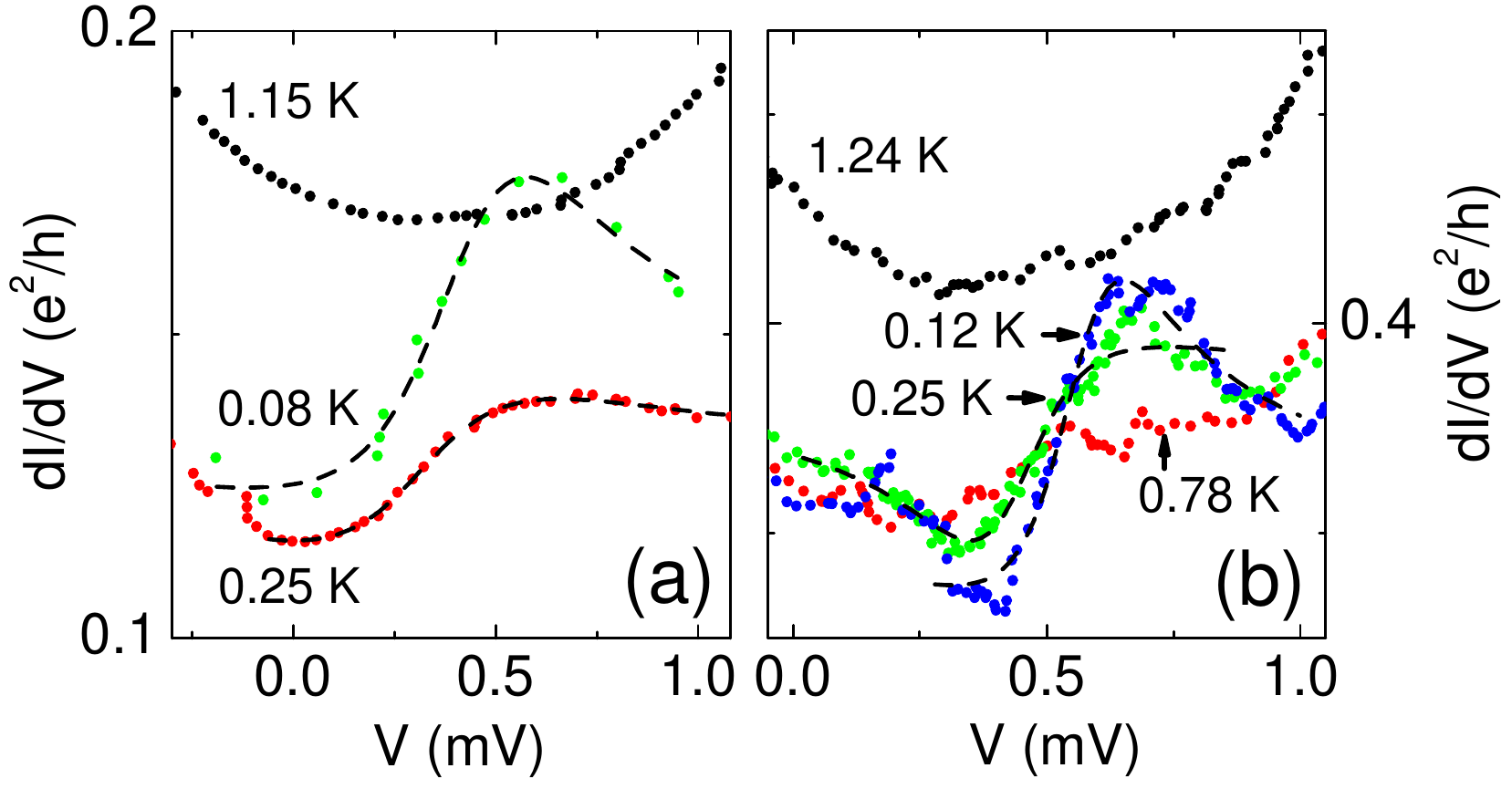}
\caption{Asymmetric resonance features of $dI/dV$
seen in two samples: 990530s1(a) and 990530s6(b). Dashed lines are
the fits of Fano's formula.}
\label{fano}
\end{figure}

By cooling down the devices to even lower temperatures, we observe
that $G$ develops a narrow resonance-like anomaly at very low bias
(Fig.\ \ref{fano}). The asymmetric anomaly builds up consistently with the
decrease of temperature and even shows a dip structure. The line
shape resembles that of a Fano resonance, which is ubiquitous in
resonant spin-flip scattering. Fano resonance happens when there
are two interfering scattering channels: a discrete energy level
and a continuum band. It has recently been rediscovered in
mesoscopic systems such as semiconductor quantum dots, SWNTs, etc.
\cite{Gores, Wiel, Nygard}. Taking a quantum dot as an
example, it can be considered as a gate-confined droplet of
electrons with localized states. The coupling of the dot to the
leads can be tuned in a controllable way to enter different
transport regimes: If the dot is weakly coupled from the
environment, a well-established CB is developed. The charge
transport is suppressed except for narrow resonances at charge
degeneracy points. When the tunnel barriers become more
transparent, the dot enters Kondo regime, i.e., below a
characteristic Kondo temperature $T_K$, spin-flip co-tunneling
events introduce a narrow symmetric TDOS peak at $E_F$ that can be
interpreted as a discrete level. If the coupling is strong enough,
the interference between this discrete level and the conduction
continuum give rise to an asymmetric Fano resonance. In
Ref.\ \cite{Gores}, such a crossover from a well-established CB
through Kondo regime to Fano regime has been clearly observed.

We found that the asymmetric resonance line shape of $G$ can be
fitted by the Fano's formula: 
\begin{equation}
G \propto (E + q)^{2}/(E^{2} + 1)
\end{equation}
Here $q$ is the so-called asymmetry parameter. $E = (\varepsilon -
\varepsilon_0)/(\Gamma/2)$ is the dimensionless detuning from
resonance. As expected, the asymmetric parameter $q$ as a measure
of the degree of coupling between the discrete state and the
continuum increases when the temperature drops. The Kondo
temperature, estimated from the relation $\Gamma = 2k_BT_k$, is
consistent for a device at different temperatures, and agrees with
the observed FWHM of the resonance. 

\begin{table}
\begin{center}
\caption{\label{tab:table2}Characteristics of the two samples in
Fig.\ \ref{fano}}
\label{table2}
\begin{threeparttable}
\begin{tabular}{cccccccc}
Sample &$T$ (K) &$q$ &$\varepsilon_0$ (meV) &$\Gamma$ (meV) &$T_K$ (K)\\
\hline
990530s6& 0.25 & 0.53 & 0.43 & 0.32 & 1.86\\
990530s6& 0.12 & 1.75 & 0.57 & 0.29 & 1.68\\
990530s1& 0.25 & 0.92 & 0.3 & 0.65 & 3.77\\
990530s1& 0.08 & 1.97 & 0.43 & 0.56 & 3.25\\
\end{tabular}
\end{threeparttable}
\end{center}
\end{table}

\begin{figure}
	\centering
		\includegraphics[width=0.70\textwidth]{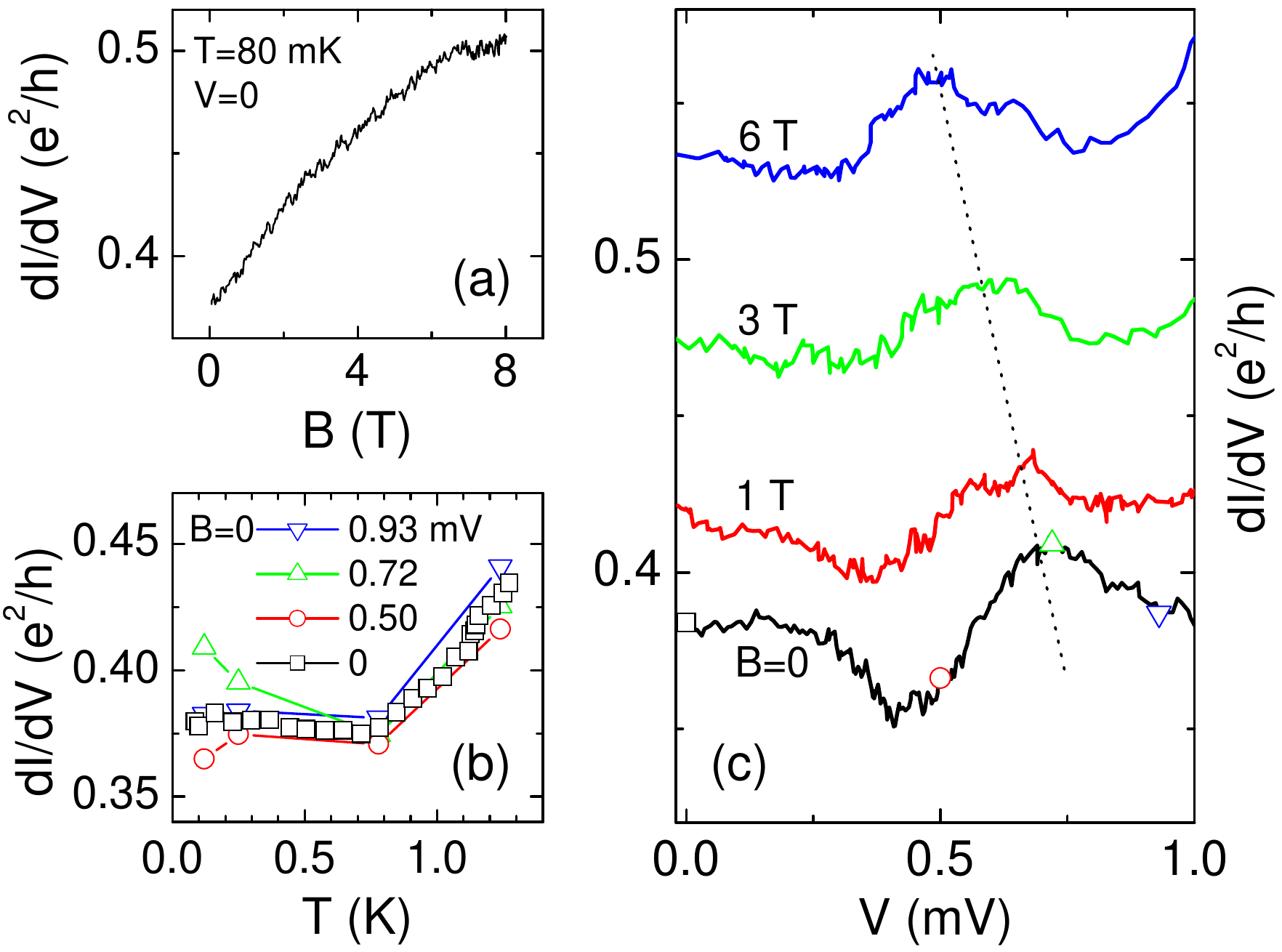}
	\caption{(a) The dependence of zero-bias $dI/dV$ on
perpendicular magnetic field. (b) The temperature dependence of
$dI/dV$ measured at different voltages. (c) Effect of
perpendicular magnetic field. From bottom to top: $B = 0, 1, 4, 8$
Tesla (sample 990530s6).}
\label{Hfield}
\end{figure}

If Kondo physics really accounts, then $G$ at voltage near resonance 
should show a non-monotonic temperature dependence around $T_K$ \cite{Nygard}. 
A close look at I--V curves (Fig.\ \ref{fano}) does exhibit such a behavior:
$G$ at the peak position first drops with $T$ and then rise up
below $\sim 1$ K. The conductance measured at zero bias, however,
shows a monotonic temperature dependence (Fig.\ \ref{Hfield}b), which is
expected when measuring at energies out of the Fano resonance.
Moreover, we found that a perpendicular magnetic field $B$ shows
two-folded effects (Fig.\ \ref{Hfield}a and \ref{Hfield}c): Firstly 
the background conductance monotonically increases with the applied $B$.
Secondly, the dip seen at zero field gradually disappears and the
resonance turns into a nearly symmetric peak at high field,
similar to what was observed in the semiconductor quantum
dots \cite{Gores}. The magnetic field effect can be explained with
the following pictures: Firstly adding a flux should change the
amplitude and/or the phase for the resonant channels therefore
breakdown the coherent backscattering and increase the forward
transmission through the channel. Secondly, the magnetic field can
destroy the interference between the resonant and nonresonant
path, transforming a resonant dip into a peak.

Since Kondo physics is historically intepreted as the interplay
between the d-orbitals of magnetic impurities and the conduction
continuum, we have to exclude the possibility that the observed
Fano resonance come from the residual Fe/Si catalyst in the MWNTs
samples. It has been shown that magnetic impurities in MWNTs cause
an enhancement of thermoelectric power \cite{Grigorian}, which is
absent in our careful TEP measurement \cite{Kang}. Furthermore, no
Fe signature in the body of MWNT bundles could be detected within
the instrumental resolution in the X-ray and TEM studies.
Therefore the observed Fano resonance should be inherent property
of MWNTs, which can be virtually treated as quantum dots strongly
coupled to the leads. We noticed that recently Buitelaar et al.
have also seen clear traces of CB and Kondo resonance in MWNT SET
devices \cite{Buitelaar}. Their observed Kondo temperature $T_K =
1.2$ K is in good agreement of our results.

\section{Summary}

In summary, strong zero-bias suppression of TDOS is observed in
MWNTs single tunnel junctions that can be well explained with the
EQF theory. The observed exponent $\alpha \approx$ 0.25--0.35 is
found to be consistent for all the samples. Reinforced by the
observation of Altshuler-Aronov-Spivak (AAS) oscillations in magnetoresistance
measurements (presented in the next chapter), we interpret MWNTs as a quasi-2D 
diffusive molecular quantum wire in which both disorder and intrinsic 
Coulomb interaction play important role in low-energy excitations. 
Similar to the case of an open quantum dot, in low-impedance junctions 
we found that a Fano-resonance like asymmetric conductance anomaly is 
built up below meV energy scales. The coexistence of abundant mesoscopic 
phenomena exhibits that MWNTs provide a good laboratory for the study of 
electron interactions in mescoscopic system.

\putbib[ch4-ref]
\end{bibunit}


\chapter{Transport Properties}

\begin{bibunit}[unsrt]

\setcounter{section}{0}

\section{Thermoelectric power}

\subsection{Introduction}

The thermoelectric power is an important transport property of solids. The basic picture of thermoelectric effect (or Seebeck effect) in a conductive solid is the carrier diffusion in the presence of a temperature gradient. The specimen is considered as a vessel filled with a gas of free electrons and/or holes, and when a temperature gradient $\Delta_{x} T=dT/dx$ is applied, conduction electrons and/or holes are thermally drifted in the --x direction, so as to setup an electric field $E_{x}$ as well as a thermoelectromotive force (TEF)

\begin{equation}
V_{x}=\int_{0}^{x}E_{x}dx\
\end{equation}

The thermoelectric power (TEP) or Seebeck coefficient is defined by

\begin{equation}
S=\frac{E_{x}}{\Delta_{x} T}=\frac{dV_{x}}{dT}
\end{equation}

In the case of a two-carrier system, electrons and holes give rise to polarizations opposite to each other, so that $S< 0$ when electrons predominate over holes in the contribution, and vice versa.

In the above discussion, the internal structure characters of materials are not taken into account. In actual solids, thermal vibrations of atoms can be described as combined elastic lattice wave, or in the quantum mechanical expression, these lattice waves are considered as phonons. In absense of electron-phonon interactions, a specimen can be looked on as a vessel of electron and phonon gases in which two constituents diffuse separately in the presence of a temperature gradient. However, in actual solids the momentum of the drifting carriers is irreversibly transfered to phonons and give rise to electrical resistance and Joule heating. In the thermoelectric effect, since the drift takes place in the same direction for both systems, phonons may transfer their momentum to carriers so as to enhance the polarization. Such a mechanism, the so-called phonon drag effect, has been manifested in a variety of solid materials.

Since carbon nanotube can be considered as wrapped-up graphite sheets, it is helpful to discuss the thermoelectric power of graphite firstly. The well crystallized graphite is a nearly compensated material, therefore the diffusive component of TEP is very low and its sign changes very sensitively, reflecting the delicate balance between electrons and holes. When the electronic structure is concerned, graphite is a semimetal in which the phonon drag effect is pronouced due to the following reasons. Firstly, the electron-phonon coupling which participates directly in the drag effect is very strong. Secondly, the wave vectors of the phonons, which play a dominant role in the drag effect, are small due to the small Fermi surface of graphite. Therefore the phonon-phonon collisions establishing equilibrium phonon distribution is not strong enough to diminish the phonon drag effect. As a result, the basal-plane TEP of graphite has a pronounced phonon drag induced dip at about 30K, at which the average wavenumber of the in-plane mode phonons is found to equal the Fermi wavenumber.

\subsection{Experimental}

\begin{figure}
	\centering
		\includegraphics[width=0.50\textwidth]{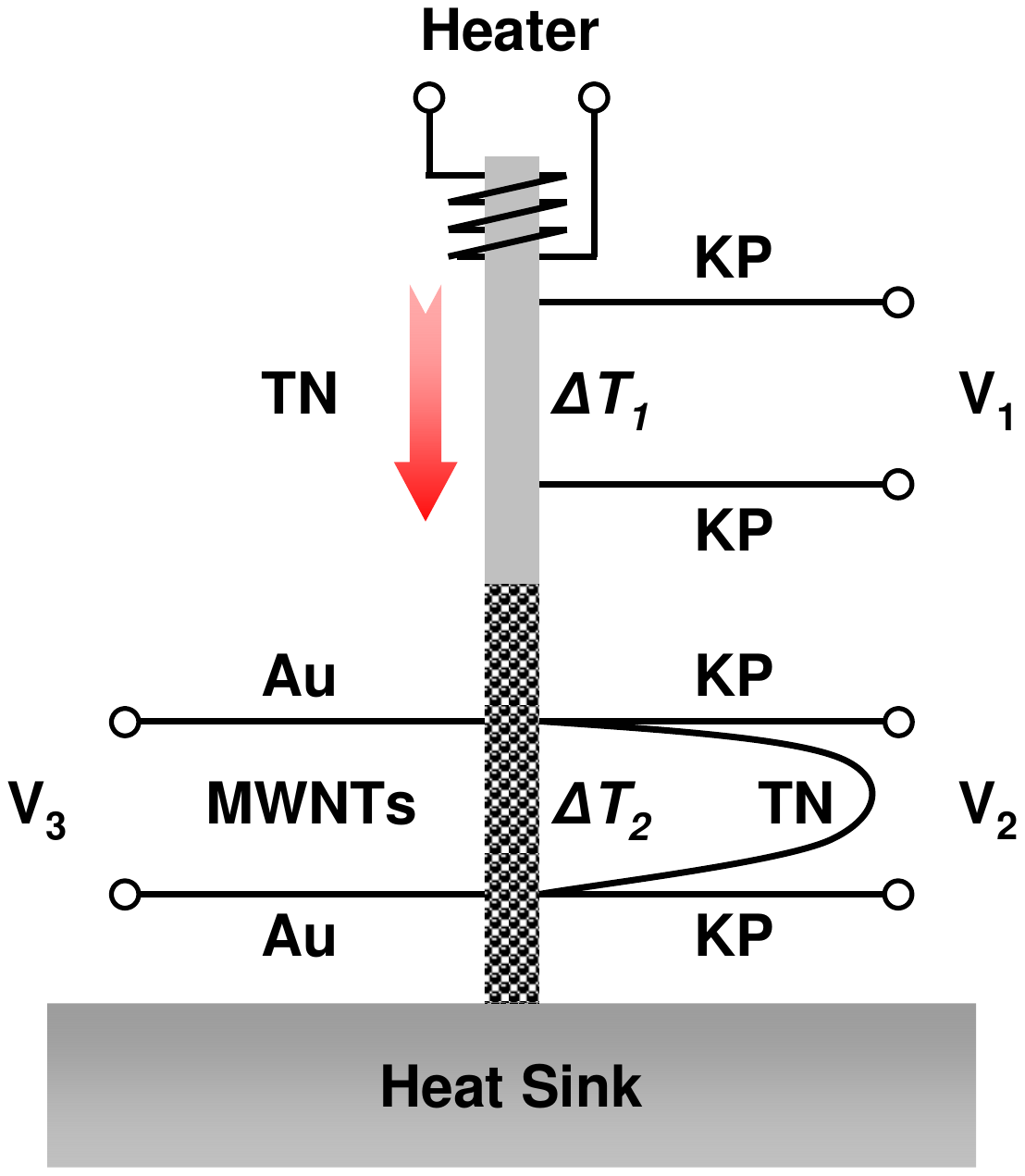}
	\caption{Schematic of experimental configuration for TEP measurement. The two Au current leads for four-probe resistance measurement are not shown.}
	\label{config}
\end{figure}

We use a standard comparative method to measure the TEP and thermal conductivity of a MWNT bundle sample simultaneously. The measurement configuration is shown in Fig.\ \ref{config}. The sample was placed in series with a standard constantan (TN) rod between the electric heater and the heat sink. Two chromel (KP) wires were attached to the constantan rod to form a thermocouple. Assuming that there were no heat losses, i.e., the same heat flow passes through the constantan rod and the sample. Four Au electrical leads were attached to the sample with silver paint so that successive measurement of electrical resistance could be performed on the same sample. Chromel-constantan thermocouple leads were attached with high thermal-conductive epoxy at the same place with the voltage leads. TEP and thermal conductance was measured at fixed temperatures by heating the far end of the constantan rod with triangle-wave formed currents at very low frequencies (i.e., 0.01 Hz). By comparing different voltages induced on the constantan rod and the sample, we obtained the thermal conductance of the tubes.

According to the formula of Seebeck effect, $S$ of two chromel-constantan thermocouples and MWNT sample are as follows

\begin{equation}
V_{1}=(S_{TN}-S_{KP})\Delta T_{1}
\end{equation}
\begin{equation}
V_{2}=(S_{TN}-S_{KP})\Delta T_{2}
\end{equation}
\begin{equation}
V_{3}=(S_{MWNT}-S_{Au})\Delta T_{2}
\end{equation}

The heat currents flow in the constantan rod and MWNT bundle could be written as

\begin{equation}
Q_{1}=-\kappa_{TN}A_{1}\Delta T_{1}/L_{1}
\end{equation}
\begin{equation}
Q_{2}=-\kappa_{MWNT}A_{2}\Delta T_{2}/L_{2}
\end{equation}

Here $A_{1}$ and $A_{2}$ are cross-sectional area of the constantan rod and MWNT bundle separately. $L_{1}$ and $L_{2}$ are their length. Assuming that there were no heat losses at the interface between the constantan rod and the MWNT sample, i.e., $Q_{1}=Q_{2}$, $S$ and $\kappa$ of MWNT sample can be derived

\begin{equation}
S_{MWNT}=S_{Au}+(S_{TN}-S_{KP})V_{3}/V_{2}
\end{equation}
\begin{equation}
\kappa_{MWNT}=\kappa_{TN}(A_{1}L_{2}/A_{2}L_{1})V_{1}/V_{2}
\end{equation}

\begin{figure}
	\centering
		\includegraphics[width=0.70\textwidth]{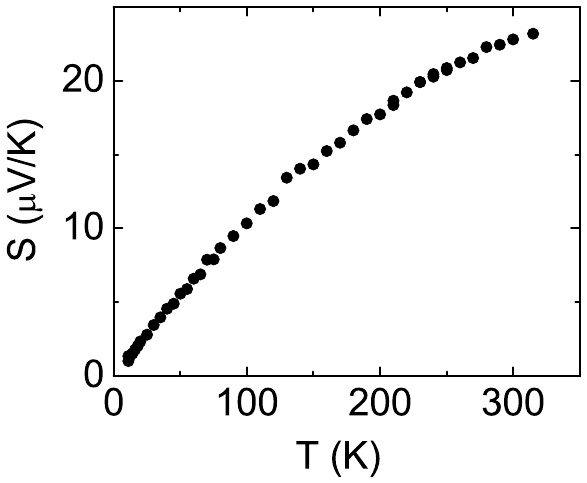}
	\caption{TEP of a MWNTs sample measured from 10 K to 300 K.}
	\label{TEP}
\end{figure}

Shown in Fig.\ \ref{TEP} are the results of our measurement. TEP of MWNTs is moderately large, approaching $\sim20$ $\mu$V/K at room temperature. The sign of TEP is positive, indicating that holes predominate over electrons in the contribution. At low temperatures, TEP has a nearly linear temperature dependence and approaches zero as $T\rightarrow0$, which is a character of metallic behaviors. At high temperatures TEP tends to be saturated. Such a large positive thermoelectric power of MWNTs gives strong evidence that the electron-hole symmetry in metallic nanotubes is broken. Similar results on SWNT ropes were reported by Hone et al. \cite{Hone}. At the present stage, the origin of such symmetry breaking is unclear. Hone et al. suggested that the electron-hole symmetry is broken due to inter-tube hopping when the tubes are assembled into ropes. However, our MWNTs are well decoupled from each other, therefore such a mechanism is not appropriate. Other mechanisms, such as carrier interactions with disorder and defects, might play important roles in TEP.

The data of thermal conductivity (not presented) is qualitatively similar to that derived by $3\omega$ method. In most of the temperature range measured, $\kappa$ is nearly linear. The order of magnitude of $\kappa$ is the same with that derived by $3\omega$ method, giving a convincing proof that $3\omega$ method is very reliable.


\section{Magnetoresistance}

\begin{figure}
	\centering
		\includegraphics[width=0.70\textwidth]{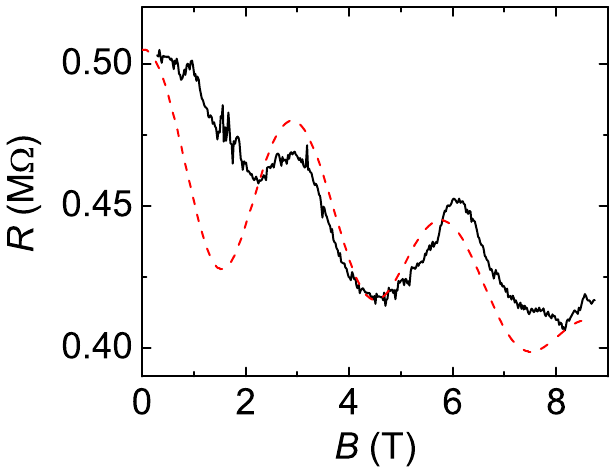}
	\caption{Longitudinal Magnetoresistance of a MWNTs sample at 20 mK (solid curve). The temporal resistance jumps of the bare data have been removed. The dashed curve corresponds to theoretical prediction for the quantum correction to the resistance of a thin-walled cylindrical conductor using the parameters mentioned in the text.}
	\label{AAS}
\end{figure}

In this section, the dependence of the electrical resistance of MWNTs on longitudinal magnetic field (parallel to tube axes) has been studied. Without access of lithographic micro-fabrication facilities, we prepared the devices by the following method: A thin bundle of MWNTs was laid on a sapphire substrate with parallel chromium electrodes separated with each other by 4 $\mu$m. Then a zircite wafer with a layer of single-sided Scotch tape was pressed on the sample by a tiny screw. The Scotch tape was used to prevent the breaking of the sample. By adjusting the magnitude of pressure, good electrical contacts could often be made. The room-temperature two-probe resistance of the samples range from several tens k$\Omega$ to 1 M$\Omega$. The experiment was conducted in a dilution refrigerator with a superconducting magnet (Oxford Instruments 200 TLM). Figure\ \ref{AAS} shows a typical longitudinal magnetoresistance (MR) measurement. When a magnetic field $B$ is applied, the resistance of MWNTs decreases. This negative MR is associated with the phenomenon of weak localization (WL). WL originates from the enhancement of backscattering which contains interference terms adding up constructively in zero field. In a specific geometry like a diffusive and thin-walled metallic cylinder (or ring), the WL contribution is periodic in the magnetic flux through the cylinder with period $h/2e$, known as Altshuler-Aronov-Spivak (AAS) effect \cite{AAS1,AAS2}. In Fig.\ \ref{AAS}, it can be clearly seen that the resistance has a second maximum at $B=3$ T and a third maximum at $B=6$ T. From TEM images, the average outermost diameter of our MWNTs is 30 nm, the corresponding period of AAS oscillations is 2.93 T, which is in good agreement with the observation. 

\begin{figure}
	\centering
		\includegraphics[width=0.70\textwidth]{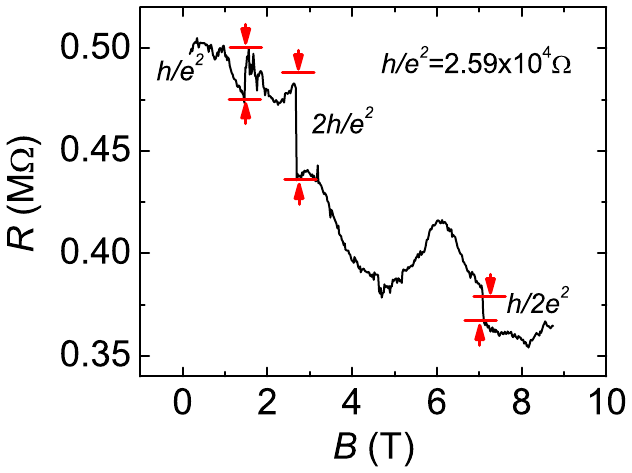}
	\caption{The raw data of the same sample as in Fig.\ \ref{AAS}. Temporal resistance jumps with magnitude $h/2e^2$, $h/e^2$ and $2h/e^2$ can be clearly seen.}
	\label{jump}
\end{figure}

Besides the reproducible resistance oscillations, non-reproducible temporal resistance jumps are often observed (Fig.\ \ref{jump}). We note that Bachtold et al. observed similar jumps in measurements on individual MWNTs. They suggested that temporal changes in the electric contacts, defects or impurities in the nanotubes are possibly responsible for these jumps. However, we note that the magnitudes of these temporal resistance jumps are integer multiples of $h/2e^2$. Therefore, the resistance jumps probably correspond to the openning or closing of the conducting channels of a MWNT. Further investigations are needed to clarify this problem.

Now let's compare our experimental data with the theory of AAS effect \cite{AAS1,Aronov}. Firstly we consider a single graphene cylinder with radius $R$ and longitudinal height $b$, which approximately equals to the distance between the electrical contacts. Then two parameters appear in the AAS theory: the temperature-dependent phase-coherence length $L_{\varphi}$ and the mean free path $l$. Following the formula given by Altshuler, Aronov, and Spivak, the correction of longitudinal DC conductance is

\begin{equation}
\Delta G=-\frac{e^2}{\pi^2\hbar}\frac{2\pi R}{b}[\ln{\frac{L_{\varphi}}{l}}+2\sum_{n=1}^{\infty}K_{0}(n \frac{2\pi R}{L_{\varphi}})
\cos(2\pi n\frac{2\phi}{\phi_0})]
\label{AASformula}
\end{equation}

where $K_0(x)$ is the modified Bessel function. The conductance correction oscillates with a period $\phi_0/2$, becoming exponentially small for $L_{\varphi}\ll 2\pi R$.

In Eq.\ \ref{AASformula} the wall thickness was not taken into account. In the real samples the field in the cylinder walls is not negligible. As a result, the oscillations decay as the field increases and a monotonic component appears in the magnetic field dependence of the sample resistivity. Moreover, the magnetic field direction may deviate from the axis of cylinder by a small angle $\theta$, making the effective wall thickness $a^{\ast}$ larger than the actual value $a$.
The inclusion of magnetic field results in the phase relaxation length $L_{\varphi}$ becoming field dependent. For a cylinder with an effective wall thickness $a^{\ast}$ in a parallel field,
\begin{equation}
\frac{1}{L_{\varphi}^2}=\frac{1}{D\tau_{\varphi}}+\frac{1}{3}(\frac{a^{\ast}eH}{\hbar})^2
\end{equation}
where $a^{\ast}$ is given by
\begin{equation}
a^{\ast 2}=a^2\cos{\theta}^2+6R^2\sin{\theta}^2
\end{equation}
Matching the theory to the measurement at $T=20$ mK, we obtain $L_{\varphi}\approx63$ nm, $l\approx1$ nm, $\theta=4.4^{\circ}$. The agreement between theory (dotted line) and experiment (solid line) is quite good except for the low field part. This agreement is obtained when assuming that only one graphene cylinder contributes to the conductance, i.e., the wall thickness $a=0.34$ nm. The estimated value of the phase relaxation length $L_{\varphi}$ is comparable with the circumference of graphene cylinder, i.e., $L_{\varphi}\approx 2\pi R$, and is close to that derived by Bachtold et al. \cite{Bachtold-Ch5} on individual MWNT at $T=0.3$ K.
Theory also points out that the phase of the oscillations, i.e., max$R$ for $H=0$, will be reversed by the presence of spin-orbit scattering. In our case, phase reversal does not happen, indicating that spin-orbit scattering is negligible.

Therefore, we conclude that quantum-interference corrections to the resistance can account for the measured longitudinal MR. The good agreement between theory and experiment give a convincing proof that the electric current flows in the outermost metallic graphene wall of MWNTs at low temperatures. 

\section{Summary}

Transport properties, including thermoelectric power (TEP) and longitudinal magnetoresistance (MR), have been investigated on millimeter-long aligned multiwall carbon nanotubes (MWNTs). A large and positive TEP with a nearly linear temperature dependence is observed, indicating that the electron-hole symmetry in metallic nanotubes is broken. The low-temperature longitudinal MR of MWNTs shows an oscillation behavior, which is attributed to Altshuler-Aronov-Spivak (AAS) effect. This gives clear evidence that quantum interference phenomena dominate the MR at low temperatures. The phase relaxation length $L_{\varphi}$ is found to be comparable with the circumference of the outmost graphene wall. The observation of resistance jumps at integer multiples of $h/2e^2$ suggests that MWNTs are mesoscopic conductors with finite conduction channels.


\putbib[ch5-ref]
\end{bibunit}


\chapter{Conclusions}

  Main conclusions from our studies on the thermal and electrical properties of MWNTs are as follows:
\vspace{0.5cm}

1. The phonon structure of a MWNT with diameter of several tens of nm is two-dimensional. The inter-wall interactions in a MWNT is very weak, so that one can treat a MWNT as a few decoupled two-dimensional (2D) single wall tubules.

2. The low thermal conductivity indicates that large amounts of defects exist in the MWNTs derived by thermal decomposition method.

3. Strong Coulomb blockade induced zero-bias suppression of tunneling density of states (TDOS) is observed in single tunnel junctions between MWNTs and a normal metal. Quantitative analysis
by Environmental Quantum Fluctuation (EQF) theory found excellent agreement with experimental data. A universal exponent $\alpha \approx$ 0.25--0.35 is attributed to strong intratube Coulomb interaction in MWNTs. In low-impedance junctions, Fano-resonance like asymmetric conductance anomaly is found below mV energy scales.

4. The thermoelectric power (TEP) of MWNTs is moderately large. At low temperatures, TEP has a metallic-like linear temperature dependence. Such a large positive thermoelectric power of MWNTs gives strong evidence that the electron-hole symmetry in metallic nanotubes is broken.

5. The longitudinal magnetoresistance (MR) of MWNTs shows an Altshuler-Aronov-Spivak (AAS) effect induced oscillation at 20 mK. This result gives clear evidence that quantum-interference dominate the resistance of MWNTs at low temperatures. The estimated value of the phase relaxation length $L_{\varphi}$ is comparable with the circumference of the graphene cylinder.



\begin{CJKAcknowledgments}
\begin{CJK*}{GB}{gbsn}
\CJKtilde
\vspace*{-.8in}
  {\centerline{\bf {\LARGE 致\  \ 谢}}}
\vspace*{.2in}
\indent
 
     本论文是在我的导师吕力研究员的悉心指导和关怀下完成的。吕
老师敏锐的物理直觉、严谨的治学态度、身先士卒的工作作风和博大
的胸怀对我产生了深刻的影响。两年的研究经历极大地开阔了我的视
野，锻炼了我的独立思考和动手能力。我取得的每一点进步都凝聚了
吕老师的艰辛劳动和大量心血。在这里谨向吕老师致以衷心的感谢。
     还要特别感谢张殿琳老师在实验方案方面给予我的直接指导和帮
助。张老师渊博的学识、活跃的学术思想使我受益非浅。在研究过程
中，物理所412组的解思深研究员、潘正伟博士不仅提供了样品，还
给予了积极的合作和大力支持，作者向他们表示感谢！
     感谢清华大学物理系范守善教授、物理所陶宏杰老师、孙刚老
师、胡占宁老师在百忙之中抽时间与我讨论实验结果并提出了宝贵的
建议。感谢清华大学物理系博士生胡辉同学和本组博士生何海丰同学
在碳纳米管的隧道谱方面的有益的讨论。
     感谢物理所502组潘明祥老师、赵德乾老师、509组赵士平老
师、王瑞峰同学和杜寰同学在实验设备和样品制备方面提供的热情帮
助。
     在硕士学习期间，本人得到了本研究组的林淑媛老师、李山林老
师、景秀年老师和王云平老师的热情指导和帮助，以及刘恩生、何海
丰、赵彦明、刘淑梅、林岚五位同学的大力支持与帮助。作者向他们
表示衷心的感谢。在物理所507组度过的时光是难忘的。
     最后，感谢我的女友陈玉两年来的理解和支持。感谢我的父母以
及所有的亲属对我的关怀和帮助！

\vskip 30pt

谨把本文献给我亲爱的父母

\end{CJK*}
\end{CJKAcknowledgments}

\newpage

\begin{acknowledgments}

This dissertation could not have been completed without the meticulous guidance and care of my supervisor, Prof. Lu, Li. Prof. Lu's sharp scientific intuition; strict work attitude; leading roles in experiment; and open mind have deeply affected me. Two years of research experience greatly broadened my horizons, tempered my ability of independent thinking and first-hand skills. Every point of my progress embodies Prof. Lu's painstaking efforts. Here I would like to express my heartfelt gratitude to him.

Special thanks to Prof. Zhang, Dianlin for his direct guidance and help on the experimental schemes. I greatly benefited from Prof. Zhang's vast knowledge and active academic thinking. In the course of my study, Prof. Xie, Sishen and Dr. Pan, Zhengwei at Group 412 not only have provided samples, but also have given enthusiastic cooperation and strong support, I am very grateful to them.

I want to thank Prof. Fan, Shoushan at Physics department, Tsinghua University; Prof. Tao, Hongjie; Sun, Gang; and Hu, Zhanning at Institute of Physics; for spending time with me despite their busy schedules to discuss the experimental results and having made valuable suggestions. I am also thankful to Ph.D. candidate Hu, Hui at Physics department, Tsinghua University; and Ph.D. candidate He, Haifeng in my group; for the helpful discussions on the tunneling spectroscopy measurements.

Thanks to Prof. Pan, Mingxiang and Zhao, Deqian at Group 502; and Prof. Zhao, Shiping; Ph.D. candidates Wang, Ruifeng and Du, Huan at Group 509 for their help in the laboratory equipments and sample preparations.

Many thanks to Dr. Lin, Shuyuan; Li, Shanlin; Jing, Xiunian; and Wang, Yunping in my group for their guidence and support throughout my M.S. study. I am also thankful to my fellow group members Liu, Ensheng; He, Haifeng; Zhao, Yanming; Liu, Shumei; and Lin, Lan for their support and help. The time I spent at Group 507, Institute of Physics is cherishable.

Finally, I would like to thank my girlfriend Chen, Yu for her understanding and support in the past two years. This dissertation is dedicated to my parents for their ubiquitous care and love throughout my life.

\end{acknowledgments}

\begin{revnotes}

This dissertation is revised from the original version submitted seven years ago. At the time I had not learned how to include figures in LaTeX. Instead, figures were hard copied and the thesis was submitted in printed format only. The main purpose of the revision is to reserve a digital copy and make it available online for interested readers. For authenticity, the main content of the thesis is kept As Is, except for the necessary corrections and reformatting. Noticeable changes are listed as follows:

\begin{enumerate}

\item	Chapter 4 (Tunneling Spectroscopy) is updated according to the recent publication (PRL 91, 076801 (2003)). The corresponding part in Abstract is updated as well.

\item In Chapter 1, Fig. 1.3(b) is changed. Original figure (curves of constant energy of graphene sheet) is changed to the current version (DOS of carbon nanotubes from STM and tight-binding calculations). Fig. 1.5 is corrected. Original figure (band structure of (5, 5) nanotubes) is replaced by band structure of (10, 10) nanotubes to make it consistent with the context.

\item In Chapter 2, two new figures (Fig. 2.1 and 2.4) are added as complementary materials. Fig. 2.3 is replaced by the published version from Ref. [3].

\item In Chapter 3, Fig. 3.3 (from Ref. [22]) is replaced by more recent data (from Ref. [17]). Fig. 3.5 and 3.9 (omitted in the original thesis) are now included.

\item In Chapter 5, Fig. 5.1 (configuration for TEP measurement) is redrawn to include more details.

\item Summary sections are added for Chapter 2 and Chapter 5.

\item English translation of Acknowledgments is added.

\end{enumerate}

A recent survey by Michael Banks at the Max Planck Institute has ranked carbon nanotubes as one of the top five topics in physics publications. Although this does not promise that someday they would replace silicon microelectronics, or make space elevators, carbon nanotubes and related materials remain attractive for both fundamental research and potential applications. Since the nanotube research is evolving very fast, the content of this thesis may become obsolete. Somehow I still feel obligated to finish the revision, not just for fun, but also to make a repository of the pleasant memories of my first scientific safari.

\vspace{1.0in}

\begin{flushright}
Wei Yi\\
Cambridge, Massachusetts\\
Aug. 2006
\end{flushright}
\end{revnotes}




\end{document}